\documentclass[10pt]{IEEEtran}
\usepackage{graphicx}
\usepackage{url}
\usepackage{complexity}
\usepackage{booktabs}
\usepackage{amsfonts}
\usepackage{amssymb}

\begin{document}

\title{Applying Formal Methods to Networking:\\\emph{Theory, Techniques and Applications}}
\author{\authorblockN{Junaid Qadir and Osman Hasan}
\authorblockA{\\School of Electrical Engineering and Computer Science (SEECS),\\
National University of Sciences and Technology (NUST), Islamabad, Pakistan\\
\emph{\{junaid.qadir,osman.hasan\}@seecs.edu.pk}}
}
\date{}
\maketitle

\section{Abstract}

Despite its great importance, modern network infrastructure is remarkable for the lack of rigor in its engineering.
The Internet which began as a research experiment was never designed to handle the users and applications it hosts today. The lack of formalization of the Internet architecture meant limited abstractions and modularity, especially for the control and management planes, thus requiring  for every new need a new protocol built from scratch. This led to an unwieldy ossified Internet architecture resistant to any attempts at formal verification, and an Internet culture where expediency and pragmatism are favored over formal correctness. Fortunately, recent work in the space of clean slate Internet design---especially, the software defined networking (SDN) paradigm---offers the Internet community another chance to develop the right kind of architecture and abstractions. This has also led to a great resurgence in interest of applying formal methods to specification, verification, and synthesis of networking protocols and applications.  In this paper, we present a self-contained tutorial of the formidable amount of work that has been done in formal methods, and present a survey of its applications to networking.

\section{Introduction}

The networking industry in a way is a victim of its own popularity. Internet, which began as a research experiment in the late 1960s, became popular before many aspects of Internet's design could be formally contemplated and designed \cite{day2007patterns}. The overwhelming success of the Internet led to the need of rapid innovations in applications and protocols. This has helped develop a culture that values engineering judgment, heuristics, and running code\footnote{The ethos of the Internet research is reflected in the famous quote of David Clark: ``We reject: kings, presidents and voting. We believe in: rough consensus and running code''.} more than it values sound engineering and rigorous verification.  Unfortunately, the expedient rapid innovations resulting from this approach has resulted in a hit-and-trial hacking based software development culture. In contrast to well-honed verification and testing tools available for other fields such as ASIC hardware design, large-scale software systems, the networking industry has a very primitive testing tool-chain. The lack of rigor in networking industry, on the other hand, can be observed by the fact that simulation based testing---which is inherently a trial-and-error process---is routinely used to `establish' the correctness of networking protocols, software, and hardware. With exponential number of possibilities, exhaustive testing is almost always impossible and thus subtle bugs remain unchecked and undetected until they manifest themselves at invariably inopportune times where the consequences of bugs in the wild can be drastic \cite{garfinkel2005history} \cite{tassey2002economic}. Such a lack of rigor is totally unacceptable in most other mature engineering or manufacturing fields, and the networking community is increasingly realizing the need for better tools and techniques for verification and testing. Using formal methods will allow us to not only verify the properties of protocols and systems, but also will help us deepen our conceptual understanding of large classes of protocols.

A standard technique to manage complexity in computer systems is to utilize abstractions and modularity. Apart from the lack of a developed verification tool-chain, the Internet also suffers from a paucity of useful abstractions, especially for the control plane, which has led to accumulation of a ``big bag of protocols'' (documented in more than 7000 RFCs!) \cite{rexford2012report}. This is in contrast with other fields of computer science: e.g., the software industry has matured to incorporate a hierarchy of abstractions designed to simplify the task of programming while ensuring correctness---e.g., in software development, the high-level end-to-end requirements are separated from the low-level machine code by various abstractions such as algorithms, programming languages, compilers, tracers and debuggers, static analysis tools, etc. The lack of abstractions has resulted in an unwieldy complex Internet architecture, with under-developed underlying principles and theoretical foundations, that is totally ill-suited to the kind of dependence that is expected of the modern Internet.

Formal methods---computer techniques based on mathematical logic---are poised to play a central role in future networking as the research community increasingly converges towards a firm realization that traditional informal methods are grossly inadequate for \emph{specification}, \emph{analysis} and \emph{validation} of networking protocols \cite{zave2011experiences}. Formal methods have been extensively applied to the \emph{verification} of hardware design \cite{kern1999formal}, communication protocols \cite{bochmann1980formal} \cite{holzmann1993design} (e.g., routing protocols \cite{bhargavan2002formal}), secure software systems \cite{jurjens2005secure}, engineering systems \cite{buede2011engineering}, programming languages \cite{godefroid1997model}, network simulations \cite{bhargavan2002verisim}, large software programs \cite{chen2004model}, etc.

Unfortunately, there has been an impression in the networking community that formal methods do not return benefits commensurate with the effort to use them. Vint Cerf has written that ``Formal methods have not yielded results commensurate with the effort to use them. They are overblown, verbose, hard to use, hard to understand.'' \cite{vintcerf}. This criticism has unfortunately resulted from the lack of appreciation of advances in formal verification and sometimes due to poor communication between the formal verification community and the networking community. It is imperative in today's world, and it will become increasingly important in the future, to move away from manual error-prone methods of verification and automate as much of the verification tasks as we can \cite{holzmann2009oopsla}. Formal methods are still useful even if they do not meet the utopian ``gold standards'' of complete automation and complete generality of mathematical proofs---in particular, interactive theorem proving, abstracted models, and light-weight methods are highly suited to certain niche applications \cite{zave2011experiences}. Advances in modern technology has fortunately facilitated development of many automatic and semi-automatic tools that can be conveniently used by practitioners with limited specialized background knowledge of formal methods.

With the increasingly central role networks play in all aspects of our lives (business, personal, entertainment, etc.), the correct functioning of networking protocols and systems has never been more important. In recent times, there has been significant interest in the application of formal methods to networking \cite{forte}, not only due to the importance of this subject, but also due to the possibilities created by recent architectural developments in the networking community. In particular, the software defined networking (SDN) architecture, which proposes splitting of the control/ data planes and the management of multiple data planes through a centralized controller to allow programming the network in a software-like fashion, makes networking accessible via formal methods. This has accentuated the networking community's interest in applying formal methods to networking \cite{summerschool}. With the use of formal methods in networking, the field of network verification looks set to evolve from the current set of ad-hoc verification tools and emerge as an engineering discipline.


\emph{Contributions of this work:} In this paper, we provide a self-contained tutorial covering the vast amount of work that has been done in the area of formal methods with a special focus on their applications in the domain of networking. Due to the great breadth of the subject, and vast amount of works in associated fields, we cannot hope to be comprehensive in every respect---nonetheless, we provide an extensive, self-contained, description of application of formal methods to networking with an adequate background on logic, programming languages, automatic verification, etc. This work is different from existing surveys \cite{babich2002formal} \cite{woodcock2009formal} \cite{zhang2012verification} \cite{lopesnetwork} in its exclusive focus on application of formal methods to networking and incorporation of new trends that have emerged with recent network architectural developments (such as the development of the SDN networking architecture). The emergence of SDN, and other recent innovations, have spurred a surge of interest in the application of formal methods to networking \cite{shenkerStanford}. Our paper is timely since, despite the recent focus and interest in our subject area, there does not exist a unified survey paper that a networking researcher can use to develop a high-level broad understanding of formal methods and techniques and learn about their applications in the context of networking. This paper attempts to fill this void, and will be valuable to networking researchers who wish to exploit the large amount of work done in the formal methods community to build reliable future networks whose correctness is formally verifiable.

The remainder of this paper is organized as follows. The necessary background on logic is provided in section \ref{sec:logicfoundation}. Various tools for specification are described in section \ref{sec:toolsforspecification}. Different methods for formal verification, such as model checking, theorem proving, static analysis, etc., are described in section \ref{sec:formalverification}. The role played by ideas in programming languages is introduced in section \ref{sec:languages}. Various applications of formal methods to networking is surveyed in section \ref{sec:applications}. Various open issues and future works are identified in section \ref{sec:openissues}. Finally, this paper is concluded in section \ref{sec:conclusions}.

\section{Logic---the Foundation of Formal Methods}
\label{sec:logicfoundation}

Logic is the branch of knowledge that focuses on systemizing truth, reasoning, and inference. Studied by generations of philosophers (Socrates, Plato, Aristotle, Kant, etc.), logic has a rich ancient tradition in philosophy \cite{scholz1961concise}. Logic was developed in ancient Greece as a device for systematizing \emph{deduction} through which true statements, or \emph{conclusions}, could be derived from \emph{premises}---statements that are \emph{assumed} to be correct. Although, utilized in mathematics at least since Euclid (2300 BC), the incorporation of logic into a mathematical framework has occurred mostly in the last two centuries  \cite{ben2012mathematical} through the efforts of Frege, Peano, and Russell to axiomatize mathematics. In the field of computer science, logic has been referred to as the ``the calculus of computer science''\footnote{In a metaphorical reference to the central place calculus occupies in natural sciences.} \cite{bradley2007calculus} to highlight its pivotal, and indeed ``unusually effective'' \cite{halpern2001unusual}, role in the fields of formal methods \cite{huth2004logic}, artificial intelligence \cite{russell2003artificial}, and theoretical computer science \cite{devlin1995logic}. Formal methods, which utilize logic for modeling and reasoning about computer systems, have been extensively for formal verification of computer systems (both hardware and software) \cite{huth2004logic}.


\emph{What is logicism:} As per Aristotle's definition, logic is new and necessary reasoning---\emph{new} since we learn what we did not know, and \emph{necessary} because the conclusions are inescapable. Leibniz dreamed of such a mechanical system of reasoning which he called \emph{calculus ratiocitinator} to calculate new and necessary conclusions from facts described in a logical symbolic language, which Leibniz called \emph{characteristica universalis}. Frege devised a set-based logical language for developing Mathematics on a solid footing. Frege (1848-1925) conceived of an ambitious project, called \emph{logicism}, which aimed at deducing mathematics (more specifically, set theory, number theory, and analysis) from laws of logic \cite{franco2009history}.  This project after Frege was taken up most notably by Russell, along with Whitehead, who embarked on an ambitious project to put mathematics on firm foundations. The use of symbolic notation, an integral component of Russell's attempt to formalize mathematics, allowed rapid progress and allowed emphasis on the structure and the form of reasoning.

\emph{The `failure' of logicism: }It was discovered by Russell that a logical language based on naive set theory---which defined sets to be a collection of objects and allowed sets to contain sets (including possibly itself) as elements---could not be used as the foundation of all mathematics because it suffered from paradoxes. Russell showed the following simple example, known as Russell's paradox, to illustrate this: does the set $S$ of all sets that do not contain itself contain the set $S$ itself? This riddle exposed that naive set theory is not sufficient to act as a foundation of mathematics leading to axiomatized set theory and various typed set theory to deal with the self-referential that created the Russell's paradox. In mathematics, the standard form of axiomatic set theory is the Zermelo-Franenkel set theory with the axiom of choice (ZFC) which acts as the most common foundation of mathematics. Eventually, Fregian logic also had to be restricted---into what is now known as first-order logic---to deal with Russell's paradox, and this restricted logic was incorporated by  ZF set theory. In 1931, Godel dealt a deathly blow to logicism when he proved that any axiomatic system capable of expressing the laws of arithmetic is \emph{incomplete}---i.e., there will always be some truth of arithmetic that cannot be proved using the axioms of the system.  While logicism `failed' in its aim of deducing arithmetic from the axioms of logic, it was instrumental in establishing the limits of computation and of ``formal reasoning''. It helped identify the limits of computation and of axiomatized logic systems.


\subsection{Components of logical reasoning}

In modern terms, every logic-based language is defined in terms of three components: syntax, semantics, and proof theory. The \emph{syntax} of a language specifies all the components that can be part of a well-formed formulae. The purpose of standardizing a syntax is to aid in understanding, communicating, and reasoning. The \emph{semantics} of a language, informally speaking, deals with the ``meaning'' of the formulae, or sentence, formatted according to the language's syntax. In logic, the semantics of a language specifies the truth of a formulae with respect to each possible world \cite{russell2003artificial}. As an example, $x + y = 2$ is true when $x$ and $y$ are both equal to 1 but false in a world where $x$ and $y$ are both equal to 2. More formally, the term `model'---which is used in the name of a technique known as ``model checking'' that we shall see later in section \ref{sec:modelchecking}---is used in logic in place of ``possible world''. The meaning of a statement $\mathcal{M}$ is a model of $\alpha$ (commonly depicted as $\mathcal{M} \models \alpha$, and read as $\mathcal{M}$ models $\alpha$) is that the formulae $\alpha$ is true in situation represented by model $\mathcal{M}$. The concept of \emph{logical entailment} is similar: we can denote in notation $\alpha \models \beta$, i.e., the formulae $\alpha$ entails the formulae $\beta$ if and only if every model in which $\alpha$ is true, $\beta$ is true as well. In other words, logical entailment $\alpha \models \beta$ implies that if $\alpha$ is true, $\beta$ must also be true. Lastly, \emph{proof theory} is concerned with manipulating formulae according to certain rules.

\subsection{Propositional Logic}
\label{sec:proplogic}

Propositional logic, also called propositional calculus or sentential logic, was developed into a formal logic by Chrysippus and developed further by the Stoics and eventually by Leibniz\footnote{Leibniz is also credited for being the developer of symbolic logic, along with his more famous contributions towards development of calculus}. Propositional logic differs from syllogistic logic, proposed by Aristotle, in that it focuses on propositions which are declarative sentences that can only take values of \emph{True} or \emph{False}. Since the propositions are akin , to Boolean variables, propositional logic is also known as Boolean logic \cite{russell2003artificial}. Propositional logic is important for two main reasons. Firstly, it is fundamentally important for computer systems since it is the theory behind digital circuits.  It is also important since more complex logics (such as first-order logic, also called predicate logic, which is covered in section \ref{sec:predlogic}) builds upon propositional logic.

In propositional logic, new propositions are generated from old through \emph{truth-functional connectives} \cite{makinson2012sets}, which define the formal grammar of propositional logic, such as the \emph{not} operator ($\lnot$), the \emph{and} operator ($\land$), the \emph{or} operator  ($\lor$), the \emph{if, or implies, or the conditional} operator ($\to$), and the \emph{iff, or equivalence, or the biconditional} operator ($\leftrightarrow$). Although, the Boolean propositional operators have intuitive analogues in natural language, they are defined formally. Sometimes, the mathematical terminology has a direct analogue with our intuition: e.g., the Boolean operator \emph{and} is an operator that is defined to give a \emph{true} value if and only if applied to two expressions whose values are true \cite{ben2012mathematical}. At other times, the mathematical terminology may extend our intuitive interpretation: e.g., mathematical usage of the implication logical connective extends the intuitive concept of implication by divorcing the concept of causality from implication \cite{devlin2003sets}. Similarly, the Boolean operator \emph{or}, when applied to two expressions, has the intuitive analogue of \emph{inclusive or}, i.e., any one or both expressions are true. It is important to stress that these operators are formally defined through a truth table, and these operators may not exactly match our everyday understanding of these words. The truth tables of the logical connectives used in propositional logic can be seen in table \ref{tab:truthtablePropositional}.

\begin{table}
\centering
\caption{Truth-table of Truth-functional connectives.}
\begin{tabular}{c c c c c c c c }
  \toprule
  $\alpha$  & $\beta$ & $\alpha \land \beta$  & $\alpha \lor \beta$ & $\alpha \to \beta$  &  $\alpha \leftrightarrow \beta$ & $\lnot \alpha$ & $\lnot \beta$\\
  \midrule
  T & T & T & T & T & T & F & F\\
  T & F & F & T & F & F & F & T\\
  F & T & F & T & T & F & T & F\\
  F & F & F & F & T & T & T & T\\
  \bottomrule
  \hline
\end{tabular}
\label{tab:truthtablePropositional}
\end{table}

\emph{Propositional logic formulae:} The formulae of a formal language built on propositional logic are expressions that can be recursively built from propositional variables by using connectives. There are four important concepts that apply to formulae. Two of these concepts are important \emph{properties} of a formulae: \emph{i)} being a tautology, \emph{ii)} being a contradiction, while the remaining two concepts refer to \emph{relations} between formulae: \emph{iii)} tautological implication, and \emph{iv)} tautological equivalence.

There are two fundamental concepts that deal with formulae of all logics: \emph{i) satisfiability}---is this formula \emph{ever} true? and \emph{ii) validity}---is this formula always true? It may be noted that the satisfiability problem is very general, indeed various computer science problems can be reduced to a satisfiability formulation. Determining the satisfiability of sentences in propositional logic was the first problem that was proved to be \NP-complete \cite{russell2003artificial}. Similarly, determining the validity of logic formulae is an extremely important problem. Another important problem that deals with propositional logic is the propositional tautology or equivalence checking. 

Traditionally, propositional logic has been regarded as uninteresting due to several limitations. While propositional logic is trivially decidable in theory, the propositional satisfiability (SAT) problem is the canonical \NP-complete problem which makes it intractable in practice. Fortunately, most practical propositional SAT problems can be solved efficiently in practice. There has been a remarkable upsurge of interest in propositional logic in the last decade or so since a diverse class of problems (including scheduling, planning, problems) can be expressed as propositional satisfiability problems.

\subsection{Predicate Logic}
\label{sec:predlogic}

Developed initially by Frege and Peirce, predicate logic enhances propositional logic---which only allowed propositional symbols along with operators---with predicates, functions, and quantifiable variables. Predicate logic expressions can include: \emph{i)} propositional symbols, \emph{ii)} predicates, \emph{iii)} functions and constant symbols, \emph{iv)} quantifiers, \emph{v)} equality, and, \emph{vi)} variables \cite{russell2003artificial}. It was felt that truth-functional connectives of propositional logic (such as \emph{not, and, or, if, iff,} etc.) alone were not rich enough to capture the much richer logical structure of natural language which often uses quantifiers, or modifiers, such as `there exists', `all', `some', `among', `only', etc. This has motivated the desire to develop a  richer, more nuanced, logic. To capture the modal quantification of every day life, predicate logic, or quantificational \cite{makinson2012sets} logic \cite{huth2004logic}, allows for a universal quantifier, $\forall$, meaning `for all', and an existential quantifier, $\exists$ meaning `for some'. Predicate logic is extremely important, especially for our subject topic of formal verification of computer systems, as it is used to formalize the semantics of programming languages, and to specify and verify programs.

Propositional logic and predicate logic are also called \emph{propositional calculus} and \emph{predicate calculus}, respectively, since both of these logics, like calculus, define a set of symbols and a system of rules for manipulating those symbols \cite{bradley2007calculus}. Propositional logic and predicate logic are calculi for reasoning about propositions and predicates, respectively. It is worth emphasizing the difference between a proposition and a predicate. A proposition is a statement that is either true or false---for example, IPv4 addresses are 32 bits long is a \emph{true} statement. A predicate, on the other hand, is used to capture relation(s) or dependence on some input parameter(s)---a predicate evaluates to true or false depending on some input parameters. In the case of a \emph{unary predicate}---e.g., $x$ is a philosopher---the truth of the statement depends on the a solitary input variable. For \emph{binary predicates}, however, the truth of a statement depends on two input variables---e.g., $x > y$ depends on the values of both $x$ and $y$. In general, predicate logic may have $n$-ary relations between objects \cite{russell2003artificial}.


\subsection{First-Order Logic}
\label{sec:firstorderlogic}

Predicate logic can be categorized into various orders depending on how the quantifiers are used in predicate logic.
In first-order logic, it is assumed that the world contains objects (such as switches, routers, users, etc.), relations (faster than, happens after, etc.), functions (one more than, next hop of, etc.), and quantifiers through which facts can be expressed about some or all the objects in the universe \cite{russell2003artificial}. In first-order logic, quantifiers can range over individuals, whereas in second-order logic, the quantifiers can also range over sets, or relations. Higher-order logic can also be defined, with $\omega$-order logic being essentially the simple theory of types. First-order predicate logic is very popular amongst mathematicians and is the language of choice for most mathematicians \cite{ben2012mathematical}. While predicate logic subsumes first-order logic, second-order logic, or infinitary logic, etc., the unqualified use of predicate logic typically refers to first-order logic. First-order logic was delineated by Hilbert, and then Skolem who proposed building set-theory on the basis of first-order logic. It has been shown that first-order logic, along with a sufficiently powerful axiom system, has sufficient expressiveness for formulating virtually all of mathematics. First-order logic, like propositional logic, is a complete system \cite{franco2009history} (first proved by Godel in his completeness theorem). There are various useful verification tools that are based on first-order logic including the Alloy analyzer (which we will discuss later in Section \ref{sec:lightweightformal}).

In 1928, David Hilbert proposed the \emph{Entscheidungsproblem}, German for the `decision problem' \cite{strichman2010decision}, which asked for an algorithm which will take a statement of a first-order logic as input, and answer if the statement is universally valid---i.e., valid in every structure satisfying the axioms---with a ``yes'' or a ``no''. Hilbert's intent was to find a system for completely axiomatizing, and formalizing, all mathematical knowledge and proofs. In 1936, Alonzo Church and Alan Turing independently showed that a general solution to the Entscheidungsproblem is impossible---thus, no mechanical, or algorithmic, method can prove the validity of arbitrary predicate logic statements. The Church-Turing result for the Entscheidungsproblem also has significant implications for the use of \emph{automatic theorem proving methods for software systems}. In particular, we cannot write a program (written in any common language such as Java, C, etc.) which will be able to always answer the decision question: given a logical formula $\phi$ in predicate logic, does $\models \phi$ hold, yes or no?

The unfortunate implication of this is that no automatic deductive verification tool can exist that will work with any arbitrary predicate logic formula instance as an input and always terminate while producing a correct `yes'---corresponding to a valid input formula---or a `no' answer corresponding to an invalid input formula \cite{huth2004logic}. This poses a fundamental, and insurmountable, problem to the automatic theorem proving approach of verification, also known as automatic deductive verification.  Therefore, first-order logic, unlike propositional logic is only a semi-decidable theory---i.e.,  there exists an effective method for telling if any arbitrary given formula is in the theory, but it may give either a negative answer or no answer at all when the formula is not in the theory.

\subsection{Higher-Order Logic}
\label{sec:highorderlogic}

A \emph{higher-order logic} (HOL) is more expressive than first-order logic as it uses some additional quantifiers along with stronger semantics. Unlike first-order logic in which variables can not denote predicates, variables in second-order logic can denote predicates allowing the logic to talk about itself more easily. There can be higher-orders beyond second-order logic.  The main strength of HOL is that it is highly expressive, and can express any mathematical theory, like multi-variable calculus \cite{harrison_13} and probability \cite{Mhamdi_2013}, in its true form. The higher expressiveness associated with higher-order logic, however, is tempered with the downside that model-theoretic properties of higher-order logic are less well-behaved than those of first-order logic. In particular, validity in higher-order logic is not even semi-decidable (or anywhere in the arithmetical hierarchy).

\subsection{Hoare Logic}
\label{sec:hoarelogic}

\emph{Hoare logic} (also known as \emph{Floyd-Hoare logic} or \emph{program logic}) is a formalism that defines logical rules---i.e., axioms and inference rules---to provide an axiomatic basis for verifying computer programming \cite{hoare1969axiomatic}. The central construct used in Hoare logic is the partial correctness specification in the form of a \emph{Hoare triple}\footnote{The Hoare triple is also known as partial correctness assertion or PCA and is partially based on Floyd's intermediate assertion method}: $\{P\}$ $C$ $\{Q\}$ where $P$ is the pre-condition, $Q$ is the post-condition, and $C$ is the command. Hoare logic builds upon other conventional logic, e.g., first-order logic, for specifying the pre- and post-conditions.

Hoare Logic is a deductive proof system for Hoare triples $\{P\}$ $C$ $\{Q\}$. The partial correctness specification $\{P\}$ $C$ $\{Q\}$ means that whenever $C$ is executed in a state satisfying $P$, and if the the execution of $C$ terminates, then the terminating state after $C$'s will satisfy $Q$. Hoare logic deals with verification of partial correctness of a command, and termination of a program has to be separately proved to show total correctness. The generality of Hoare's approach is based on its characterization of programming constructs as transformations of states which can universally apply to any imperative programming language construct. The underlying semantics of a program can be viewed a set of transformations from an initial state to a final state. Since a sequential program can also be envisioned as a transformational system, Hoare logic is particularly suited to analysis and verification of sequential computer programs.  Hoare logic is a sound system (every provable formula is true) but not a complete system (i.e., not all true statements are provable). More details about Hoare logic can be found in \cite{apt1981ten}.

Hoare logic, and the use of Hoare-style pre-conditions and post-conditions, is commonly used in many settings. As an example, the Java Modeling Language (JML) defines a specification language for Java programs, following the design by contract paradigm, which uses Hoare style pre-conditions and post-conditions and invariants for extended static checking. The same style is inherited by ESC/ Java.

\subsection{Modal Logic}
\label{sec:modallogic}

Modal logic is an expressive form of logic that uses additional quantifiers. Modal logic was originally developed by philosophers to study different `modes of truth'---e.g., an assertion $P$ may be false in the present world, however, the assertion `possibly $P$' will be true if the assertion $P$ is true in some alternate world \cite{emerson1990temporal}. Temporal logics essentially have two kinds of operators: logical operators (loaned from traditional the logic framework in which temporal logic is used) and modal operators. The modal operators capture in modal logic the intuitive notions of \emph{necessarily, always, possibly, sometimes}, etc. The symbols $N, F, G, A, E$ represent \emph{\textbf{N}ext}, \emph{\textbf{F}uture}, \emph{\textbf{G}lobally}, \emph{\textbf{A}ll} and \emph{\textbf{E}xists}, respectively. In typical notational terms, the box symbol is used to represent \emph{necessity}, while the diamond symbol is used to represent \emph{possibility}. For example, $Gp$ would mean always $p$; $Fp$ will mean sometimes $p$; $\diamond p$ means possibly $p$; $\square p$ means necessarily $p$.

Modal claims can be understood semantically in a theory of ``possible worlds''---an idea commonly attributed to Leibniz which was advanced by Saul Kripke in the late 1950s. Kripke advanced Leibniz's conception of the actual world being one ``possible world'' amongst other, by proposed a mathematical theory of models (now known as Kripke models) for possible worlds. A statement is ``possible'' in modal logic if it is true in at least one possible world; a statement is ``necessary'' if it is true in all possible worlds. We will see later that ``model checking'' (covered in section \ref{sec:modelchecking}) depends fundamentally on the concept of possible worlds and utilizes Kripke models.

\subsection{Temporal Logic}
\label{sec:temporallogic}

The use of temporal logic, a special type of modal logic, for formal specification and verification of computer systems was proposed by Amir Pnueli in a highly influential paper \cite{pnueli1977temporal} in 1977. In this paper, Pnueli argued that temporal logic---a formalism for dealing with how truth values of assertions change over time---is especially appropriate for describing reactive systems such as operating systems and network communication protocols. In a reactive system, which contrast with sequential terminating programs that essentially transform the input to the output and then terminate, the normal behavior is to engage in a nonterminating computation that continuously interacts with the environment. Examples of reactive systems include operating systems and network communication protocols. Temporal logic is especially invaluable in the field of model checking finite-state \emph{concurrent} programs \cite{clarke1986automatic}: Leslie Lamport, in his highly cited paper ``what good is temporal logic?'', has highlighted that the main utility of temporal logic is in modeling concurrent systems \cite{lamport1983good}.

Temporal logic formulae differ from ordinary Boolean formula in that the temporal formulae have new modal operators---which allow qualitative description of temporal events by implicitly incorporating temporal ordering of events---in addition to the traditional Boolean operators---``and'', ``or'', ``not'',  and ``implies'' \cite{gabbay2000temporal} \cite{manna1992temporal}. The usage of temporal logic has been widely adopted for use with finite-state programs with algorithmic methods available that can verify the temporal-logic properties of finite-state systems. While the capacity to only include finite states may appear too limiting, it turns out that a wide range of systems, especially, hardware systems and communication protocols, can be modeled as finite-state programs. Some (linear) temporal logic operators include $G$ (\emph{Globally}), $F$ (\emph{Eventually, Finally}), $X$ (\emph{Next}), and $U$ (\emph{Until}). For example, we may want to reason about the temporal properties of a protocol in the following way: a message is not received unless one is sent, a message that is sent is eventually received, etc.

Temporal logic has been extensively applied to computer systems, and is a key component of the popular model checking approach (discussed in section \ref{sec:modelchecking}), because it can capture two keys notions of computer performance. Firstly, temporal logic can capture ``\emph{liveness}'' property that some good thing will happen in the future---i.e., the form $Fp$, which indicates that some proposition will be true in the \emph{future} in the course of the computation. Secondly, the ``\emph{safety}'' property of the form $Gp$ can capture the desire that globally $p$ is ensured which incorporates the proposition that undesirable states are never obtained. In addition, the ``\emph{fairness}'' property is also defined which states given certain conditions, an event will occur, or will fail to occur, infinitely often. The fairness property is often expressed with $Gp$ (infinitely often) and $Fp$ (eventually always). Efficient methods exist that can work with temporal logics. While validity in first-order logic is semi-decidable (i.e., it is possible that complete proof procedures will run forever on invalid formulas), validity/satisfiability in many temporal logics is decidable. 

There are two important subtypes of temporal logic: linear temporal logic (e.g., LTL)---where each moment in time has a unique future trajectory or possible future---and branching temporal logic \cite{lamport1994temporal} in which each moment can be split into many different possible futures. \emph{Linear temporal logic (LTL)} is a susbset of the more complex CTL that additionally allows branching time and quantifiers. LTL is also sometimes called propositional temporal logic, abbreviated PTL. LTL can use both propositional and first-order forms. LTL is popularly used, in both these forms, in the specification and verification of programs \cite{emerson1990temporal}. The SPIN model checker \cite{holzmann1997model} is based on LTL and has been extensively used for communication protocol verification \cite{holzmann1991design}. \emph{Computation tree logic} (CTL) is an example of branching temporal logic that has additional path quantifiers such as $A$ (for all paths $\forall)$ and $A$ (there exists a path) that denote universal and existential quantification over paths starting in a certain state. CTL is used mostly for applications in hardware verification, while LTL is used mostly for applications in software verification. While CTL and LTL do have overlapping expressiveness, each logic can express properties outside the domain of the other---e.g., LTL can express fairness properties which CTL cannot, but CTL can express the so-called reset property which LTL cannot. The NuSMV model checking tool is based on CTL. CTL is extensively used in the formal verification of reactive networked systems. As an example, it is used in the recent work of Reitblatt et al. \cite{reitblatt2012abstractions} which also uses model checking with the NuSMV \cite{cimatti2002nusmv} tool for verification. Other works that incorporate CTL include the ConfigChecker tool \cite{al2009network}, the Splendid Isolation project \cite{gutz2012splendid}, etc. Lastly, we will mention that the  \emph{computation tree logic star} (CTL$\ast$) logic, not as commonly used as LTL and CTL, has been proposed as a generalization of both LTL and CTL.

\subsection{Other Logics}

\vspace{2mm}
\emph{Relational Logic:} The logic used in the Alloy analyzer \cite{jackson2006software} is a relational logic that combines the quantifiers of first-order logic with the operators of the relational calculus. Relational logic extends first-order logic by incorporating transitive closure allowing greater expressiveness  Since first-order logic is undecidable, the focus of the Alloy analyzer is in \emph{model finding} rather that exhaustive model checking---in particular, not finding a model does not preclude a model in a larger scope. Most tools for relational notation, other than Alloy analyzer, e.g., PVS etc., focus instead on theorem proving and are thus not fully automated. Kodkod  \cite{torlak2007kodkod} is an example tool that is based on the relational logic of Alloy. The inclusion of ``transitive closure'' enables expressiveness (beyond that offered by first-order logic) that can be used to encode common reachability constraints. Since the relational logic of Alloy uses multi-arity relations instead of functions over sets, it is first-order and thus amenable to automatic analysis due to its simplicity.

\vspace{2mm}
\emph{Router Logic} : Feamster et al. \cite{feamster2003towards} proposed \emph{routing logic} to define a set of rules  that can be used to determine if a routing protocol satisfies various properties. Feamster et al. also utilized this logic for analyzing the behavior of BGP protocol under various conditions. Importantly, Feamster et al. suggested that in addition to analysis of existing configurations, router logic can be used to synthesize network-wide router configurations from a high-level description.

\subsection{Satisfiability of logic formulae: the SAT problem}

A fundamental concept that applies to all logic formulae is the concept of \emph{satisfiability}: is this formula ever true? The Boolean satisfiability (abbreviated as SAT) problem is an important problem in theoretic computer science having wide-range applications. The SAT problem can be defined as:  Given any arbitrary formula, find a satisfying assignment or prove that no satisfying assignment is possible. Such an assignment may not always exist---in which case, we will say that the problem is over-constrained, and the solver will report that satisfying the formula is not possible. The Boolean satisfiability problem is also alternatively known as propositional satisfiability or simply as the satisfiability problem. The SAT problem was the first problem shown to be \NP-complete\footnote{Any instance of \NP-complete problem can be transformed into an instance of another \NP-complete problem quite easily. As an example, both graph coloring and SAT problems are \NP-complete, and an instance of the former problem (i.e., graph coloring) can easily be transformed into an instance of the latter (i.e., SAT).}, and many practical problems can be reduced to a SAT formulation \cite{biere2009handbook} and solved through off-the-shelf SAT solvers.


The SAT problem has applications in scheduling, automated theorem proving, planning, model checking, software verification, synthesizing consistent network configurations, etc. SAT solvers are thus very versatile tools useful for solving constraint satisfaction problem in a variety of settings. The SAT problem is at the very heart of the problems of design, specification and verification of computer systems \cite{huth2004logic} for diverse logics. The problem of formal verification fundamentally deals with the satisfiability relation expressed as $\mathcal{M} \models \phi$ where $\mathcal{M}$ is a \emph{model} of a system and $\phi$ is a specification expressing what should be true in situation $\mathcal{M}$.


\emph{What is satisfiability mathematically?} A logic language is composed of logical symbols with fixed interpretation (e.g., in propositional logic, the logical connective such as $\wedge$, $\lor$, etc. are logical symbols) and other non-logical ones (such as propositional variables $p$, $q$, etc.) whose interpretations may vary. These symbols can be combined together to form \emph{well-formed logical formulae}. A formula is \emph{satisfiable} if it has an interpretation that makes it logically true. In this case, we say the interpretation is a \emph{model}; a formula is \emph{unsatisfiable} if it does not have any model. A logical formula is valid if it is logically true in any interpretation. Conversely, a propositional formula is valid \emph{if and only if} its negation is unsatisfiable. As an example, consider a Boolean variable $p$. The formula $p \wedge \lnot p$ is unsatisfiable since it is not true in any interpretation---in other words, it does not have any model. The formulas $p$ and $\lnot p$ are, on the other hand, satisfiable but not valid since they are true in some, but not all, interpretation(s). Finally, the formula $p \lor \lnot p$ is valid since it is true in all interpretations. In the SAT problem, we seek a \emph{satisfying assignment} for a given propositional formula on a set of Boolean variables which assigns values to the variables such that the formula evaluates to \emph{True}.

\vspace{2mm}
\subsubsection{Variations of SAT}

While we are mostly interested in propositional satisfiability due to its tractability, the concept of satisfiability can be generalized to other Boolean logics---in particular, the quantified Boolean formulas (QBF) problem generalizes the SAT problem\footnote{In the SAT problem, all the variables are implicitly existentially quantified.} and refers to the problem of deciding the satisfiability of quantified Boolean formulae, or QBF, in which the variables can be either universally or existentially quantified. The ability to utilize universal and existential quantifiers in arbitrary ways makes QBF considerably expressive than SAT. It must be noted that SAT is NP-complete which means any NP problem can be encoded in SAT. Similarly, QBF is \PSPACE-complete, i.e., any PSPACE problem can be encoded in QBF. Unfortunately, current QBF solvers do not scale, and therefore, our primary focus will be on the SAT problem and solvers.

The SAT problem has many interesting variations. For example, the MaxSAT problem is the application of SAT problem to optimization theory, the AllSAT problem aims to determine all satisfying assignments, etc. Motivated by the success of SAT solvers, researchers have recently given significant attention to Satisfiability Modulo Theories (SMT). In the SAT problem, the logical operatives were restricted to the conjunctive normal form (CNF) and qualifiers such as ``for all such things'', or ``there is one such thing'' were not allowed. The SMT problem is considered more difficult than the SAT problem \cite{malik2009boolean}. While SAT solvers determine the satisfiability of propositional formulas, SMT solvers can, on the other hand, check the satisfiability of formulas in some decidable first-order theory (e.g., linear arithmetic, array theory, uninterpreted functions, bit-vectors, etc.) \cite{de2011satisfiability}. SMT is seeing rapid progress and initial commercial use in software verification \cite{barrett2009satisfiability}.

\vspace{2mm}
\subsubsection{SAT/ SMT solvers}

Since the SAT problem is \NP-complete, the general problem is theoretically intractable. All currently known SAT  solutions thus perform poorly in the worst-case---i.e., with exponentially increasing computation cost as the instance size increases. Fortunately, the intractability of the general SAT  problem does not practically rule out efficient solutions of special cases. There has been great advances recently in the field of formal verification based on the discovery that SAT solvers can solve a wide variety of practical SAT problems quite efficiently \cite{malik2009boolean}. Modern tools can solve practical industrial SAT problems having millions of variables and constraints in mere seconds. In practice, such approaches can help avoid the daunting proposition of redeveloping algorithmic solutions for solving new problems, thus enabling a wide variety of application areas to benefit.

Broadly speaking, there are \emph{two ways to use a SAT solver}. The first, and simplest way, is the \emph{eager approach} for the application to generate a Boolean formula for the SAT solver so that it may determine that the satisfiability of the formula. Alternatively, the application can use the \emph{lazy approach} to reduce a problem to a series of inter-related SAT queries, in which the SAT solver incrementally solves subsequent queries dynamically generated based on the results of previous queries \cite{bordeaux2006propositional}. Much of the improvement in SAT solver performance in recent years has been driven by several improvements to the basic DPLL algorithm such as \emph{i)} non-chronological backjumping and learning conflict clauses; \emph{ii)} optimization of ‘constraint propagation’ rules; \emph{iii)} heuristics for picking ‘split’ variables (even restarting with a different split sequence); \emph{iv)} Highly efficient data structures. A detailed account of various algorithms for solving the SAT problem is presented in \cite{gu1999algorithms}, whereas recent advances in SAT-based formal verification can be viewed at \cite{prasad2005survey} \cite{ganai2007sat}. A comparison of propositional satisfiability and the related field of constraint programming can be seen in \cite{bordeaux2006propositional}.

Various SAT/ SMT tools have been proposed with rapid progress in this field being sustained by Moore's law and consistent advances in algorithms, data structures, and decision heuristics \cite{gomes2008satisfiability}. Example SAT/ SMT solver tools include MiniSAT \cite{een2005minisat}, Chaff \cite{moskewicz2001chaff}, and the Z3 tool from Microsoft \cite{de2008z3}. Due to the great generality of SAT/ SMT solvers, it is remarkable that various contemporary verification tools that differ in terms of source language, methodology, and degree of automation, eventually fall back on these solvers for the core task of checking validity and satisfiability. With their impressive generality, scalability, and maturity, SAT/ SMT solvers look set to play a significant role in future formal verification technology.

\vspace{2mm}
\subsubsection{Applications of SAT/ SMT solvers to Networking:}

Recent advances in SAT/ SMT solvers have significantly advanced the state of the art in formal verification, and SAT/ SMT tools are routinely used in network verification projects. We present a few works as examples. Zhang et al. have presented an approach for verifying and synthesis of firewalls using SAT and QBF \cite{zhang2012icnp}. FLOVER, a model checking system, implemented using the Yices SMT solver \cite{dutertre2006yices}, verifies that the network’s security policy is not violated by the aggregate of flow policies instantiated within an OpenFlow network \cite{sonmodel}. Recently, there has been work in verifying the data plane through SAT solvers. Anteater \cite{mai2011debugging} verifies the data plane by translating connectivity invariants into SAT problems that are checked against the data plane by a general SAT solver to return a counter example in case of violation of invariants. NetSAT is another data plane verification project that is SAT based \cite{zhang2013sat}. Some more examples of the use of SAT/ SMT technology in the context of networking can be seen in table \ref{tab:networking}.

\subsection{Algebra and Logic}

An algebra is a structure that consists of sets and operations that act on those sets. Using the tools of algebra, logical statements can incorporate unknowns, symbols, and formulas. This symbolic calculus enables correct reasoning with economy of mental effort and has led to rapid development in mathematical knowledge. To paraphrase Alfred Whitehead, symbolism facilitates understanding, and tracking of transitions in, reasoning almost mechanically by the eye without undue taxing of the brain. Mathematical logic, or symbolic logic, improved upon the logic of Aristotle by exploiting symbolic manipulations---or, essentially the methods of algebra.

\vspace{1mm}
\emph{Boolean Algebra}: The algebra of logic was founded by George Boole (1815 to 1864), and perfected by later logicians, to formalize the ``laws of thought''. Boolean algebra is essentially the `algebraization' of classical propositional logic and the bridge between logic and algebra.

\vspace{1mm}
\emph{Relational Algebra}: The field of databases extensively utilizes ideas from relational algebra.  Relational algebra, an off-shoot hybrid of first-order logic and of algebra of sets, essentially deals with manipulations of relations. The formalism of relational algebra, proposed by E.F. Codd in the 1970s, can be used as a query language for relations and serves as a theoretical foundation of databases.

\vspace{1mm}
\emph{Kleene Algebra:} The study of semantics and logics of programs utilizes Kleene algebra which defines algebraic structures with operators +, ., *, 0, and 1 satisfying certain axioms.  Kleene algebras arise in many diverse contexts: relational algebra, semantics and logics of programs, etc. Kleene algebra was extended to incorporate tests to produce \emph{Kleene algebra with tests} (KAT) \cite{kozen1997kleene}. KAT has recently been used in the NetKat \cite{anderson2013netkat} project to provide consistent reasoning principles about network applications in the setting of SDN.

\vspace{1mm}
\emph{Algebraic path-finding:} In networking context, algebra can be viewed as a concise language useful for describing combinatorial problems. Researchers have applied algebraic ideas to network routing through algebraic path finding methods that exploit the fact that numerous practical network problems are in fact instances of the same abstract ``algebraic path problem'' (e.g., a classical example of an abstract algebraic path problem is shortest path routing) \cite{baras2010path}. Routing algebra meta-language (RAML), which builds upon Sobrinho's \emph{Routing algebra} \cite{sobrinho2005algebraic}, was proposed by Griffin in the ``metarouting'' project \cite{griffin2005metarouting}. Metarouting aims at equipping network operators with the ability to define their own routing protocols in a high-level declarative manner using a domain-specific language customized for specification, verification, and implementation of routing path metrics. Sobrinho's Routing algebra \cite{sobrinho2005algebraic}, which can be understood as generalization of shortest path routing, is expressive enough to adequately model complex policy-based routing typified by ubiquitous the Border Gateway Protocol (BGP) routing protocol. A key feature of the metalanguage proposed for metarouting, which is especially relevant to our subject topic, is that algebraic properties required for guaranteeing correctness can be automatically derived.

\section{Tools for Specification and Modeling}
\label{sec:toolsforspecification}

There are three important components of a verification framework. Firstly, since it is often cumbersome and unwieldy to work with real systems, there has to be a \emph{i)} \emph{framework for modelling systems}: this typically employs a \emph{description language} of some sort---especially, when considering hardware systems. Secondly, a \emph{specification language}---typically a logic-based language---is needed for specifying the desired properties that are to be verified. Lastly, a \emph{verification method} is needed to establish if the system model satisfies the specification.

In this section, we will study techniques for modeling systems in section \ref{sec:modelingsystems} and for specifying properties in section \ref{sec:specification}. We will cover verification methods later in section \ref{sec:formalverification}.

\subsection{Modeling Systems}
\label{sec:modelingsystems}

Systems can be divided into two broad classes. \emph{Transformational systems} may be modeled as black boxes that take certain input and produce a final result as output and terminate. Such systems can modeled in terms of their input/ output relations. Formal methods developed for such transformational systems include the Floyd-Hoare logic (section \ref{sec:hoarelogic}), which allow reasoning about such systems through pre- and post-conditions, and specification languages like Z (which we will cover in section \ref{sec:specification}). \emph{Reactive systems}, on the other hand, maintain an ongoing interaction with their environment, and thus such systems must be specified and verified in terms of their ongoing behaviors. Formal methods proposed for such reactive systems have to use more sophisticated techniques than those provided by the pre- and post-conditions in notations such as Z. In particular, label transition systems (called Kripke structure) based on the concept of finite state machines (FSMs) and temporal logic have been proposed for modeling reactive systems.

In the following subsections, we will discuss various approaches for modeling systems. We will cover FSMs, Kripke structures, binary decision diagrams (BDDs), and model extraction from code in sections \ref{sec:fsm}, \ref{sec:kripke}, \ref{sec:bdd}, and \ref{sec:modelextraction}, respectively.

\vspace{2mm}
\subsubsection{Finite State Machines}
\label{sec:fsm}

The mathematical formalism of finite state machine (FSM), or finite state automaton, is commonly used in the study of the design of computer programs and sequential circuits \cite{lee1996principles}. An FSM can be conceived as an abstract machine having finite states in which the machine can be in only one state at any given time. The FSM can make a transition---i.e., change its state from the \emph{current state} to another state when triggered by some event or condition. A given FSM is defined by its set of states, and the triggering conditions for each transition. The ``state transition model'' of FSM has been extensively used in formal verification and serves as the basis of system modeling in ``model checking''.  The state transition model is amenable to mechanical automated verification, but suffers from the ``state space explosion'' problem, which describes the case when the number states of the system model becomes so large that it becomes  infeasible to exhaustively explore the state-space using the available computational resources.

Broadly speaking, there are two kinds of FSMs: \emph{i)} the more general \emph{Mealy machines}, in which the output depends not only on the system state but also on the system input, and \emph{ii)} \emph{Moore machines}, which are special cases of Mealy machines, in which the output is determined by only the system state. A FSM is deterministic if the next state and the output are uniquely determined by the current state and input, otherwise, the FSM is non-deterministic if a given state and input can non-deterministically lead to one of many possible next states and outputs. Non-deterministic FSM (NFSM) can be viewed as a generalization of deterministic FSM.

A protocol specification can be translated into a FSM model, with each asynchronous process coded as a separate FSM, extended by message queues and variables if necessary. The system remains finite and amenable to exhaustive search if the queue size and the range of variables is bounded. The system is non-deterministic in general since in each system state a number of transitions may be simultaneously executable. There are two important structural properties of FSMs when used to represent protocols \cite{holzmann1988improved}. Firstly, the state space is sparse, i.e., the set of effectively reachable state is much less than the number of potentially reachable states with a ratio of $1$ in $10^9$ being typical. Secondly, the state space is tightly connected, i.e., the states are usually reachable by mildly different paths that differ only in the order in which the execution of the asynchronous protocol is interleaved.

There is a well-developed theory for verification of FSMs: e.g., reachable states, and equivalence, etc., that can readily exploited for network verification tasks. In particular, reachability of states is very relevant in a networking context. The FSM formalism has been extensively used in formal verification works for networking \cite{reitblatt2012abstractions} \cite{al2009network} \cite{kazemian2013real} \cite{al2010flowchecker}---in these works, the packet is considered as an FSM. Many network verification projects model the network as a large state machine (see description in \cite{lopesnetwork} and \cite{zhang2012verification}). Unfortunately, the FSM verification problem is \PSPACE-complete, and therefore is computationally very complex. The problem, however, reduces to be \NP-complete if the FSM can be formulated as a combinational logic network.


\vspace{2mm}
\emph{Automata Theory} is a field of theoretical computer science that has been used in the study of computability and languages \cite{hopcroft2008introduction}. Finite automata constitute an important formalism in theoretical computer science. It is useful for modeling a wide variety of systems that have finite number of states (e.g., communication protocols, for lexical analysis as used in compilers, for scanning text, for expression pattern matching, etc.). An automaton can be envisioned as a special case of Moore machines in which only two outputs---ACCEPT and REJECT---are defined. Variations on the general theme of automata, with varying degrees of expressiveness, have been proposed \cite{lee1996principles}: e.g., timed-automata \cite{alur1999timed}, Petri nets \cite{murata1989petri}, etc. These formalisms have been adopted in the field of formal verification: e.g., Petri nets have been commonly used for representing concurrent network protocols \cite{billington1999application} while timed-automatons have been used for verifying timing properties of network protocols and real-time systems in time-automata based model checking tools (to be discussed later in section \ref{sec:modelchecking}) such as UPPAAL \cite{larsen1997uppaal}.

\vspace{2mm}
\subsubsection{Kripke Structure}
\label{sec:kripke}

Kripke structure is a labeled state transition graph that can adequately capture the temporal behavior of reactive systems. From a practical point of view, the Kripke structure is nothing but a \emph{labeled} FSM extended to incorporate a labeling function that maps states to sets of atomic propositions making it possible to specify simple propositional properties on the FSM. When used in conjunction with some temporal operators, these propositional properties can be used to specify properties like ``from a state labeled \emph{REQ}, the state labeled \emph{ACK} will eventually be reached'' \cite{wang2006abstraction}. Kripke structure can easily model diverse kinds of systems that are described using formulae of first-order logic.

Kripke structures are often used to model reactive systems that interact with the environment in a continuous fashion without terminating \cite{clarke1999model}. Since such systems do not terminate, input-output transformation characterization is not sufficient. Instead, it is important to capture the \emph{state} of the system, and how the system state changes as a result of some action. One way of doing this is by identifying the transition of the system---which describes the system state before an action occurs and after it occurs, respectively.

More formally, Kripke structures consist of a set of states, set of transitions between states,  and labels for each states defining properties that are true in that state. A Kripke structure $\mathcal{M}$ over AP, representing a set of atomic propositions, is a 4-tuple  $\mathcal{M} = (S, S_0, R, L)$ where \emph{i)} $S$ is the \emph{finite} set of states, \emph{ii)} $S_0$ is the set of initial states, \emph{iii)} $R$ is the transition relation, and \emph{iv)} $L$ is a labeling function that labels every state with the set of atomic propositions that are true in that state.

\vspace{2mm}
\subsubsection{Binary decision diagrams (BDDs)}
\label{sec:bdd}

The concept of ``binary decision diagrams'' (BDDs) is quite old but was popularized by Bryant in 1992 \cite{bryant1992symbolic} as an efficient method for representing state transition systems \cite{andersen1997introduction} \cite{knuth2009art}. It has been pointed out earlier that techniques like model checking suffer from the problem of state explosion which is quite likely to occur if the system under study is composed of components that can perform transitions in parallel. This can cause the system states to grow exponentially leading many experts to be skeptic about the ability of model checking to scale to large systems.  Model checking owes most of its success to the development of the data structure of BDDs which allows efficient verification of large transition systems. Computer science luminary Don Knuth cites BDDs as one of the most fundamental data structure development in the last 25 years which allows solutions to problems previously imagined as intractable \cite{knuth2009art}. The BDD data structure allows concise representation of large transition systems and easy manipulation, and is therefore an important component of many logic synthesis and formal verification systems \cite{knuth2009art} \cite{bryant1995binary}.

Bryant also observed that reduced ordered BDDs (OBDDs) are a canonical representation of Boolean functions. The use of reduction and ordering is common in BDDs, and in fact, the term BDD is commonly understood to refer to reduced ordered BDDs \cite{knuth2009art}. BDDs are able to reduce the space required for storing state transition systems by identifying redundancies through the following three rules: \emph{i)} merge equivalent leaves, \emph{ii)}  merge isomorphic nodes, and lastly \emph{iii)} to eliminate redundant tests.

It was noted in \cite{mcgeer2012new} that the rulesets that network administrators typically write lead to small BDDs. BDD is a very popular data structure that can be used, along with efficient graph algorithms for BDDs, to significantly improve the computing time and space efficiency of algorithms \cite{yangreal} \cite{al2009network}.

\vspace{2mm}
\subsubsection{Model extraction from code}
\label{sec:modelextraction}

One of the hindrances in the popularization of formal verification is the tediousness of the task of creating system models. A possible solution to this problem for the specific case of \emph{software systems} is to apply verification methods not to models of code, but to implementation code directly through some automated model extraction technique. Some example efforts in this domain include extension of the SPIN model checker for support of embedded software in abstract models \cite{holzmann2000automating}, formal verification of device driver code at Microsoft \cite{ball2001slam} through automatic predicate abstraction of C programs \cite{ball2001automatic}, and CMC tool at Stanford that works directly with C code \cite{musuvathi2002cmc}.

\subsection{Formal Specification}
\label{sec:specification}

In networking protocols, it is important that protocols are defined unambiguously. Traditionally, the specification process adopted by Internet Engineering Task Force (IETF) is based largely on specification through informal English prose, with implementations also serving as an informal specification surrogate. Although in the early 1980s, various IETF standards have been formally specified by various academics (including an Estelle \cite{budkowski1987introduction} description of Transport Control Protocol, TCP), the IETF has not embraced the use of formal description techniques and continues to specify protocols informally relying primarily on the implementation as the specification. The tendency to use the implementation as the specification has the drawback of not cleanly separating what is part of the protocol and must be conformed to and what is system and implementation dependent. The lack of the emphasis on formal specifications for Internet protocols has created a problem where it is considered acceptable to create software without fully understanding the implications leading to an ad-hoc hit-and-trial based software development culture \cite{jacky1996way}. Experience with Internet protocols has shown that simple informal English prose is insufficient for specifying and communicating protocols \cite{zave2011experiences}. Many of the problems that arise due to informal specifications can be redressed through formal methods for specification which aid not only in verification and communication, but also in analysis \cite{day2007patterns}. In particular, analytical tools can analyze the formal description to ensure that absence of protocol deadlock, data loss, races, hazards, and other pathological behaviors.

Formal specification can be used by the formal verification process to verify that the desired properties are held by the system model. For the purpose of formal verification, \emph{equivalence checking} can be used to match an implementation against a full specification of what a program must do. However, due to the significant overhead involved in writing a full specification, formal verification is often done with partial specification that describes only some desired behavior of the program. This endeavor which contrasts with equivalence checking is known as \emph{property checking}. Most property checking tools use either logical deductive interference or model checking, and report a counterexample when a property violation is seen. It is worth emphasizing that correctness is not an unqualified concept since correctness measures the relation between two entities: a specification and  an implementation, or a property and a  design \cite{camurati1988formal}. Thus verification is only as good as the specification, making specification an extremely important part of verification. 

Broadly speaking, \emph{formal specification techniques} can be categorized into three types based on the underlying formalism. Firstly, in the \emph{mathematical or language-based techniques}, commonly a predicate calculus based approach is taken to represent protocols. Secondly, in the \emph{FSM-based techniques}, an existing programming language may be extended to incorporate the representation of a state machine and associated rules. Techniques like extended FSMs, Petri nets, abstract state machines fall under this category. Lastly, in the \emph{temporal logic techniques}, which are especially useful for reactive systems, in which the protocol is described in terms of statements that implicitly incorporate the relative ordering of events and their actions. IEEE's ``property specification language'' (PSL) (IEEE 1850 standard) is an example specification language rooted in temporal logic that is commonly used in hardware design where it is a common practice to augment design with assertions serving to specify correct behavior.

There are many standard \emph{formal description languages} for protocols \cite{babich2002formal}. The Estelle language \cite{budkowski1987introduction} and the SDL language \cite{belina1989ccitt}, specified by CCITT/ ITU, are based on a extended state model. The LOTOS language \cite{bolognesi1987introduction}, on the other hand, is based on a temporal logic model.  The Z (pronounced Zed) language \cite{jacky1996way} is a popular formal specification language useful for describing \emph{transformational} systems such as sequential programs in Hoare style using pre- and post-conditions. PROMELA is a specification language used for specifying LTL formulas that can be used for validation of \emph{reactive} systems with the SPIN model checker. The interested reader is referred to a tutorial article \cite{babich2002formal} for more details about formal description and specification techniques such as SDL, Estelle, PROMELA, LOTOS, etc.


\section{Techniques for Formal Verification}
\label{sec:formalverification}

Various approaches have been proposed for formal verification which include both automated and interactive techniques. We discuss model checking as an example automated method in section \ref{sec:modelchecking}. We will discuss theorem proving---a technique that can automated for decidable logics such as propositional or first-order logic, but which works in concert with a human expert for dealing with the undecidable higher-order logic as a proof-assistant---in section \ref{sec:theoremprovers}. In the later part of this section, we will discuss light-weight formal methods, static analysis, and symbolic execution \& simulation in sections \ref{sec:lightweightformal}, \ref{sec:staticanalysis}, and \ref{sec:symbolicexecution}, respectively.


\subsection{Model Checking}
\label{sec:modelchecking}

Developed independently in 1980's by Clarke and Emerson\footnote{For a historical account of the development of model written, the interested reader is referred to \cite{grumberg200825} and \cite{emerson2008beginning} (written by Emerson from his personal perspective)} \cite{clarke1986automatic}, and by Queille and Sifakis \cite{henzinger1992symbolic}, model checking can be envisioned as an automated debugging, or exhaustive simulation and testing, technique useful for checking any property violations (i.e., bugs or errors) \cite{baier2008principles}. While formal verification has traditionally been associated with logic-based axiomatic or deductive techniques for establishing proofs of correctness, model checking has been the first step towards engineerization of this field \cite{baier2008principles} \cite{clarke1999model}.

The main insight of model checking is that proof construction---a tedious and non-trivial task requiring good deal of ingenuity and guidance from the user---is not necessary for the case of finite state concurrent systems. In proof-based verification, we are interested in showing $\Gamma \vdash \phi$ where $\Gamma$ is a \emph{set of formulas} representing the system description in a suitable logic, and $\phi$ is another formula representing the specification. We are interested in a deductive proof $\Gamma \vdash \phi$. Given a logical proof system that is sound and complete, $\Gamma \vdash \phi$ holds iff $\Gamma \models \phi$ (semantic entailment). Semantic entailment is undecidable for first-order logic while model checking is decidable. In model checking, we are interested in showing that $\mathcal{M} \models \phi$ where $\mathcal{M}$ represents a Kripke structure\footnote{A Kripke structure, proposed by Saul Kripke, is a nondeterministic automaton representing a system's behavior. Kripke structures are commonly used in model checking for interpreting temporal logics.}, or a labeled transition graph, as a model of system description  while the specification is still a formula (typically written in propositional temporal logic). More specifically, the model checking problem is (from \cite{clarke2008birth}): ``Let $\mathcal{M}$ be a Kripke structure (i.e., a state transition graph), $\phi$ be a formula of temporal logic (i.e., the specification). Find all states $s$ of $\mathcal{M}$, such that $\mathcal{M}, s \models \phi$ (i.e., $\mathcal{M}$ has property $\phi$ at that state $s$)''. As discussed earlier in section \ref{sec:kripke}, Kripke structures are \emph{labeled} FSM with the states labeled with a sets of atomic propositions that are true in this case; all other unlabeled propositions are assumed false according per the ``closed-world'' assumption. This model checking can be performed  for finite state systems algorithmically, unlike proof systems, in a push-button fashion. In model checking, the verification procedure intelligently searches through the entire state space of the design in an exhaustive fashion \cite{clarke1986automatic}, and thus this technique  is applicable for finite state systems\footnote{Infinite state can only be analysed with abstraction \cite{wang2006abstraction} and induction.}. Although this looks limiting, many interesting systems (e.g., hardware devices, communication protocols, etc.) can be modeled as FSMs in practice.


It is important to ensure that the term ``model'' in ``model checking'' is not confused with its everyday usage of being an abstraction of the actual system under study. In the case of ``model checking'', the inventors of this method were interested in the model-theoretic interpretation \cite{modeltheory} \cite{marker2002model} of the term `model'---i.e., determining that $\mathcal{M}$, representing the system interpreted as an automaton, is a (Kripke) \emph{model} for the temporal logic formula $\phi$ representing the desired property \cite{clarke2008birth}. It should be noted that when we say that $\mathcal{M}$ is a model for the formula $\phi$, we really are paraphrasing our intention of saying `$\phi$, when interpreted as in $\mathcal{M}$, is true'. Noting the distinction between the various interpretations of models can alleviate any unnecessary confusions. To summarize, model checkers are named such because they check whether a system, interpreted as an automaton, is a (Kripke) model of a property expressed as a temporal logic formula.

Model checking has many benefits over deductive proof techniques which makes it preferable wherever it is applicable. Some compelling benefits of model checking \cite{clarke2008birth} include: \emph{i)} it is fast compared to other rigorous methods, \emph{ii)} it provides diagnostic counterexamples, \emph{iii)} it can work well with partial specifications/ properties, \emph{iv)} logics can easily express various concurrency properties, and finally, \emph{v)} it does involve any human-guided proofs.

Buchi automata has been used in model checking as a bridge between automata theory and temporal logic. In particular, Buchi automata can provide an automata-theoretic formalization of a linear temporal logic, or LTL, formula. It was shown in the mid 1980s that there exists for every temporal logic formula a Buchi automaton that accepts precisely those runs that satisfy the formula. There are algorithms that can mechanically convert any temporal logic
formulae into the equivalent Buchi automaton. Typically, the property invariants are expressed as LTL formulas, and a negated version is converted to Buchi automata to be used in the model checking algorithm to detect violation of the desired property.

\emph{Scalability of model checking:} The state explosion problem limits the application of model checking to large scale problems. Various approaches have been proposed for coping with this issue including symbolic model checking, bounded model checking, and statistical model checking. These approaches are covered next.

\vspace{2mm}
\emph{Symbolic Model Checking:}
\vspace{2mm}

The main insight of symbolic model checking is that it is more efficient to consider large number of states simultaneously at a single step instead of traversing enumerated reachable states one at a time. Symbolic model checking facilitates such a state space traversal by allowing representations of states set and transition relations as Boolean encoded formulas, BDDs, or related data structures. This allows handling of much larger designs containing hundreds of state variables \cite{mcmillan1993symbolic} \cite{clarke1996symbolic} \cite{burch1992symbolic}. Symbolic algorithms can thus work with the FSM represented implicitly as a formula in quantified propositional logic without the need of explicitly building a FSM graph. In summary, a symbolic model checking method is a model checking method that represents state sets symbolically, typically using OBDDs, as opposed to an explicit enumeration of states. Symbolic model checking is the most commonly used variant of model checking used by most industrial scale model checking tools. The first symbolic model checking tool, SMV, was developed by McMillan in 1992 and used BDDs to combat the state explosion problem \cite{mcmillan1992smv}. More recently, SMV has been extended and reimplemented as NuSMV and NuSMV2 \cite{cimatti2002nusmv}.

\vspace{2mm}
\emph{Bounded Model Checking:}
\vspace{2mm}

Symbol model checking can also be performed through SAT procedures \cite{biere1999symbolic}. SAT procedures can operate on Boolean expressions without requiring canonical forms and without the potential space explosion of BDDs. Various efficient implementations are available for solving SAT problems. Bounded model checking (BMC) uses a SAT procedure instead of BDDs \cite{clarke2001bounded}. A Boolean formula is constructed that is satisfiable \emph{iff} there is a counterexample of length $k$. By incrementing the bound $k$, longer counterexamples can be  searched. If after some number of iterations, we may conclude that no counterexample exists and the specification holds. The state explosion problem is thus handled by focusing on falsification rather than exploring all reachable states. Incorporation of the falsification approach into a SAT based framework in a BMC allows scaling to much larger number of states. BMC techniques using the falsification approach are very useful since in many practical scenarios, we are more interested in finding bugs as early as possible in the design rather than in formally proving the correctness of the design. SAT-based BMC for falsification is a very popular model checking technique in the industry. As an example, safety property may be verified by increasing the number of iterations to the bound defined by the diameter of the FSM. The advantage of the bounded model checking approach is that it can quickly find counterexamples due to the depth first nature of SAT search procedures. Secondly, since the bound is increased incrementally, the approach finds the counterexample of minimum length which leads to better diagnostics. Finally, it also uses lesser space as compared to BDD-based approaches. The NuSMV2 tool \cite{cimatti2002nusmv} incorporates both BDD-based and SAT-based model checking. BMC can also be performed using SMT tools \cite{armando2009bounded}. BMC tools include a CBMC \cite{clarke2004tool} which is a bounded model checker for ANSI-C and C++ programs.


\vspace{2mm}
\emph{Statistical Model Checking:}
\vspace{2mm}

Statistical model checking is a proposal that can allow model checking to scale to large systems by relaxing the requirements of formal correctness. The key insight is to use hypothesis testing with a simulation based approach to deduce from some sample executions if the system under test satisfies the specification \cite{legay2010statistical}.

\vspace{2mm}
\emph{Probabilistic Model Checking:}
\vspace{2mm}

Various approaches have been proposed for building probabilistic model checking tools \cite{fagin1994reasoning} \cite{baier2003model} \cite{kwiatkowska2007stochastic}. PRISM is an example probabilistic model checking tool that can be used for reachability analysis \cite{kwiatkowska2009prism} and protocol verification \cite{kwiatkowska2002probabilistic}. While traditionally, establishing performance evaluation and correctness have been orthogonal tasks, a promising new direction in formal methods research is to develop probabilistic methods that can allow joint analysis of both correctness and performance \cite{alurtheory}.


\vspace{2mm}
\emph{Model checking for Software:}
\vspace{2mm}

Model checking is not inherently well suited for verifying software due to the asynchronous and unstructured nature of software. While, the early successes of model checking were mainly in hardware verification, recent progress has made model checking viable for software verification \cite{jhala2009software}. Popular model checking softwares include Java Pathfinder \cite{havelund2000model}, Microsoft's Slam Toolkit \cite{ball2001slam}, UC Berkeley's BLAST \cite{henzinger2003software}. The interested reader is referred to a detailed survey on model checking for software for more details \cite{jhala2009software}.

\vspace{2mm}
\emph{Applications of Model Checking to Networking:}
\vspace{2mm}

There are a great number of \emph{model checking tools} that have been devised with some popular model checkers being SPIN \cite{holzmann1997model}, NuSMV \cite{cimatti2002nusmv} and Alloy \cite{jackson2006software}. SPIN, developed in early 1980s by Holzmann for assuring dependability in complex telephone switching systems, is a popular award-winning\footnote{The SPIN model checker has been awarded the ACM Software System Award \url{http://www.acm.org/announcements/ss_2001.html}.} \emph{explicit-state} model checking tool. SPIN was the first model checker developed, with its initial focus being on telecommunication systems and protocol verification. SPIN is now used for diverse applications from hardware verification to distributed control software used in nuclear power plants and spacecrafts. The IEEE Futurebus cache coherence protocol is the first IEEE protocol whose specification was debugged successfully through model checking. NuSMV, in contrast to SPIN, is a \emph{symbolic} model checking tool that also incorporates features of \emph{bounded} model checking. NuSMV was the first implementation of symbolic model checking and was developed by McMillan in 1992 \cite{mcmillan1992smv}. NuSMV can utilize both BDD-based and SAT-based techniques. Alloy is also a symbolic model checker that translates constraints into Boolean formulas which are then solved through an external SAT-solver. SPIN and NuSMV support temporal logic for property specifications with SPIN supporting propositional LTL and NuSMV supporting CTL. For model specification, SPIN uses the PROMELA language (which is inspired by C) while NuSMV uses the SMV description language to specify finite state machines. Alloy uses first-order logic for both model specification and property specification. A detailed comparison of SPIN, NuSMV, and Alloy, and some other model checking tools, is presented in \cite{frappier2010comparison}. Popular model checking tools are listed in table \ref{tab:tools}, along with other popular formal verification tools, for quick reference. Apart from SPIN, NuSMV, and Alloy, it is worthwhile to mention two other popular types of model checking tools. The PRISM tool \cite{kwiatkowska2009prism} is a probabilistic model checking tool, while the UPPAAL tool \cite{larsen1997uppaal} is a model checking tool based on timed-automata which can be used for verification of real-time systems.

Model checking techniques and tools have been extensively applied in the context of networking, and we will present a representative sample. Zave et al. have used model checking to understand SIP \cite{zave2008understanding}. Al-Shaer et al. have used model checking for configuration analysis for general networks \cite{al2009network} and for SDN networks having federated OpenFlow infrastructures \cite{al2010flowchecker}. In \cite{al2010flowchecker}, network routing tables are represented as BDDs and reachability predicates are computed using model checking. In other works for OpenFlow networks, Canini et al. present the model checking based NICE platform for verification \cite{canini2012nice}, and Son et al. present a model checking based security invariant property checker \cite{sonmodel}. Most of the model checking work has focused on verifying safety property since verifying liveness property entails computing an infinite long trace of states in which the desired property is never reached with heuristics-based MaceMC being a notable exception \cite{killian2007life}. A summary of various applications of model checking techniques in the context of networking is presented in table \ref{tab:networking}.


\subsection{Theorem Provers}
\label{sec:theoremprovers}

In the theorem proving paradigm of formal verification, the relationship (\emph{implication} or \emph{equivalence}) between the specification and the implementation is considered as a proof goal, which is verified using a computer-based tool called a theorem prover. The dream of having automated theorem provers is a long-standing dream of many an ambitious scientists starting from Leibniz, to Peano and Hilbert \cite{chang1973symbolic}. Herbrand in 1930 provided a mechanical method for proving theorems but due to lack of appropriate computing facilities the method was difficult to apply. In 1936 Church and Turing showed that it is impossible to devise a generic method of verifying the validity of first-order logic. First-order logic is said to be semi-decidable in that methods exist for verifying validity of a formula if it is indeed valid, however, such methods will never terminate in general for invalid formulae. This has defined the limits of automatic theorem proving. In the 1960s, Herbrand's method was implemented on a digital computer, followed by an even more efficient Davis-Putnam-Logemann-Loveland (DPLL) algorithm \cite{davis1962machine}. The resolution principle, proposed by Robinson in 1965, has been a major step forward. The DPLL algorithm is important for many applications including automated theorem proving and satisfiability modulo theories (SMT). The DPLL algorithm is used to solve the CNF-SAT problem---i.e., determine the satisfiability of propositional logic formulae in conjunctive normal form (CNF).

Despite the theoretic complexity of automated reasoning in expressive logics, in practice, \emph{interactive theorem provers}---also known as proof assistants---which solve the proof verification problem are useful in many settings. Interactive theorem provers differ from automatic theorem proving in that it requires human assistance. A proof assistant is a program that takes a formalized mathematical statement and a plausible proof, and checks whether the proof is valid. There are three key ingredients of a proof assistant. Firstly, it needs to incorporate an expressive formal language and logic---which is typically, but not always, a variant of higher-order logic. Secondly, it needs to have support for checking proofs and in aiding proof construction. Lastly, it needs to have a programming language---typically a functional programming language---that allows extending the system with new proof procedures (e.g., decision procedures). There are various interactive theorem provers that have been proposed including tools that are based on first-order logic (e.g., ACL2 \cite{brock1996acl2},  Microsoft's Z3 tool \cite{de2008z3}---which uses many-sorted first-order logic, etc.) and others that are based on higher-order logic (e.g., Isabelle \cite{paulson2002isabelle}, HOL \cite{gordon1987hol}, PVS \cite{owre1992pvs}, Coq \cite{bertot2004interactive}, etc.).

\vspace{2mm}
\emph{Applications to networking:} Theorem provers have many applications in networking research. As specific examples, we will discuss three theorem proving tools that are popularly used in networking research. The Coq tool, which incorporates higher-order logic along with richly-typed functional programming language, defines a system for manipulating and mechanical verification of formal mathematical proofs by machines \cite{bertot2004interactive}. Coq also supports extracting certified programs to popular functional languages like OCaml, Haskell, etc. The Coq tool has been used for verifying the network controller in SDN environments \cite{guhaformal} and for ensuring per-packet and per-flow consistency of network updates \cite{reitblatt2012abstractions}. The Z3 tool from Microsoft, which uses a portfolio of solvers, is another popular theorem prover used in many software testing, analysis and verification projects \cite{de2008z3}. Finally, the Isabelle/HOL theorem prover  has been used for network verification and the BGP policy verification \cite{paulson2002isabelle} is a notable example in this regard.

Besides the functional verification, theorem provers have also been used for the formal performance analysis of network applications based on the higher-order-logic formalizations of probability theory \cite{Mhamdi_2013} and Markov Chains \cite{liu_2013}. Some notable examples in this regard include the performance analysis of the Stop-and-Wait protocol \cite{hasan_09}, scheduling algorithms of Wireless sensor networks \cite{elleuch_11}, the memory contention problem in multiprocessor systems \cite{liu_2013b} and the quantitative analysis of information flow in a network \cite{tarek_12}.


\subsection{Light-weight Formal Methods}
\label{sec:lightweightformal}

``Full-blown'' formal methods, such as model checkers and proof systems, have some limitations due to which there is interest in alternative ``light-weight'' formal methods \cite{agerholm1999lightweight}. Proof systems, like theorem provers, have the deficiency that they cannot be fully automated due to fundamental limits of computation. Model checkers, on the other hand, are not inherently suited to software systems due to the state explosion problem, and since they cannot deal with indirection which is a fundamental concept of software \cite{jackson2006software}. To avoid the overhead of full-blown formal methods, fully automated analysis methods based on lightweight formal methods \cite{jackson2001lightweight} have been proposed that exploit advances in technologies such as SAT solvers.

The \emph{Alloy analyzer} works by translating constraints to be solved from Alloy into Boolean constraints which are then fed to an off-the-shelf SAT solver. Alloy is also known as a \emph{model-finding} tool since it aims to find an instance of a counterexample, known as a model in logic theory, quickly rather than for completeness. Alloy defines both a language for describing structures and also a tool for exploring them---in particular, it specifies a new high-level language, inspired from Z \cite{jacky1996way}, for specifying the structure an behavior of software; secondly, it uses an automated SAT-solver based analyzer to work through all the possible scenarios. The software design modeled with a high-level notation is then analyzed over billions of possible executions to catch any pathological conditions. The important consequence is that subtle design errors are caught even before the design is coded. The design, once thoroughly tested, can then be constructed with much more confidence. Alloy true to its style of being a light-weight formal method works on a analyze-first-then-prove principle. Alloy represents a new generation of software analysis engines similar in principle to tools traditionally used for verification of hardware designs \cite{jackson2006dependable}.

\vspace{2mm}
\emph{Applications to networking:} Light-weight formal methods in general, and particularly the Alloy tool, have been widely deployed to solve a wide variety of problems ranging from security analysis \cite{narain2005network} to the design of telephone switching networks \cite{zave2011experiences} \cite{zave2012formal}.

\subsection{Static Analysis}
\label{sec:staticanalysis}

Static analysis is a class of techniques concerned with extracting information about the run-time behavior of a program, or a configuration file, without actually executing the source file. Static analysis which means analysis without execution (e.g., SLAM \cite{ball2001slam}) is to be contrasted with dynamic analysis which involves executing the program (e.g., Verisoft \cite{godefroid1997model}). Static analysis can discover bugs in configuration files, or software systems, before they are activated or executed thus obviating the reflexive debugging that results from discovery of bugs after deployment. Due to the fundamental limits imposed by the theory of computation (cf. Turing's halting problem which is notorious for being undecidable), static analyzers \emph{cannot} extract run-time behavior of all programs perfectly. Static analyzer attempt to defy the undecidability of the halting problem by not focusing on \emph{completeness} or \emph{soundness} but instead on quick and efficient debugging. The key insight used by static analysis is to utilize an approximate interpretation, or an abstract interpretation, of the program. In many cases, this approximate interpretation is finite, and thus amenable to analysis.

The term soundness has a background in mathematical logic where a system is said to be sound if it can only prove valid arguments with respect to a semantics . In the context of debugging, soundness means the ability to detect \emph{all} possible errors of a certain class, or not miss a bug if one exists---in other words, a sound debugger will give no \emph{false negatives}. Completeness, in contrast, implies that there will be no \emph{false positives} which requires exhaustive analysis of every possible scenario. An effective static analyzer, thus, has to balance three desirable, but often competing, costs: \emph{i)} the cost of false negatives due to being unsound, \emph{ii)} the computational cost of analysis, and \emph{iii)} the usability of the tool (which can be measured in total time investment of the user)  \cite{xie2005soundness}. In particular, it turns out that soundness and completeness have a tradeoff. For \emph{assurance} based projects where soundness is needed (i.e., if told that there are no errors, we should be sure that there are none), we are limited to accepting incompleteness, or to accept false alarms or false positives. The presence of false alarms is usually irritating for customers of debugging tools---who aim incidentally to reduce the number of bugs and not necessarily eliminate all of the bugs---who often give up on soundness to reduce the number of false alarms. Most commercial debugging tools (such as Coverity, etc.) are neither sound nor complete, but perform well in practice catching many errors with lesser number of false alarms.

It is instructive to compare static analysis with model checking directly \cite{engler2004static}. In general, model checking has some benefits that are hard for static analysis to match: e.g., \emph{i)} it can check the implication of code, and not just surface-visible properties, \emph{ii)} it gives stronger correctness results, etc. A major drawback of model checking approach is the need to create a correct working environment model---this restrictions makes model checking infeasible for many networking verification tasks \cite{feamster2005detecting} and adds significant overhead even when feasible especially for large scale systems. Also, no model is as good as the implementation itself, and the abstraction in the modeling process is a potential for producing false positives or missing critical errors. Static analysis is more useful than model checking in some aspects: e.g., \emph{i)} it is quicker, \emph{ii)} it can easily check millions of lines of code, \emph{iii)} it can find thousands of errors. Some of these comparisons are direct outcomes of the fact that static analysis does not run any code, while model checking does \cite{engler2004static}. Static analysis is a widely used technique used in many software testing tools (e.g. Coverity, FindBugs, etc.) that can analyze extremely large code-bases \cite{bessey2010few}.

Static analysis is familiar to all programmers in its most basic form of typechecking in compilers (e.g., a Java compiler will catch errors such as adding a number to a Boolean, etc.) This kind of static analysis focuses on simple checking with no false alarms and thus only scratches at the surface of what can be achieved with static analysis. More extensive static analysis requires more computation but can check a wider range of properties---e.g., runtime exceptions due to division by zero, array bounds violation, etc. can be detected. Since such analysis is difficult to do precisely, such extensive static analysis can involve false positives (non-errors reported as errors) and false negatives (non-reported errors).

Extended static checking defines a powerful paradigm for program checkers in which verification conditions---i.e., a logical formula that is valid iff the program is free of the classes of error under consideration---are defined, and then counterexamples to the verification condition are searched mechanically \cite{leino2001extended}. Extended static checking for Java (ESC/Java) \cite{flanagan2002extended} is a compile time program checker that performs formal verification of properties of Java source code through theorem proving.  ESC/Java provides an annotation language, which is effectively a subset of Java Modeling Language (JML), which a programmer can use to add Hoare-style preconditions and postconditions and loop invariants into the program with special comments in the source code.

\vspace{2mm}
\emph{Applications to networking:} Static analysis has also been used in networking context most notably for reachability analysis in IP networks \cite{xie2005static}, firewall analysis (e.g., Margrave \cite{nelson2010margrave}, etc.), BGP configuration fault detection \cite{feamster2005detecting} \cite{qie2004using}, etc. It can also be used for debugging of networking software using techniques described above.

\subsection{Symbolic Simulation and Execution}
\label{sec:symbolicexecution}

Symbolic execution \cite{cadar2013symbolic}, also called symbolic evaluation, is a `abstract interpretation' method for analyzing a program assuming symbolic values for inputs rather that actual inputs that would arise through the normal execution of the program. Symbol execution is essentially a technique for generating an optimized test suite that satisfies a customizable coverage criteria using which deep errors in software applications may be identified. Although the idea of symbolic execution is quite old (proposed by King in 1976 \cite{king1976symbolic}), symbolic execution has emerged as an effective tool recently with advanced in constraint satisfaction tools. Symbolic execution proceeds by exploring as many program paths as it can in a given time budget, thereby creating a logical formula encoding the explored paths. A constraint solver is then used to calculate feasible execution paths. Symbolic execution are much more powerful than dynamic execution techniques, such as those incorporated in popular debugging tools like Valgrind \cite{nethercote2007valgrind}, since it  can find a bug if there exists \emph{any} buggy input on a path without depending on a concrete input that triggers the bug. Symbolic simulation \cite{bryant1991symbolic} is an extension of the idea of symbolic execution to hardware systems. Simulation is a time-test tool for formal verification. Simulation can be generalized in two different ways: \emph{i)} ternary simulation \cite{bryant1991formal}---where we have a ``don't care'' value X in addition to 0 and 1; \emph{ii)} symbolic simulation---where Boolean variables can act as input parameters and outputs are functions of these parameters. Ternary symbolic simulation unfortunately suffers from the problem of large growth in the state space leading researchers to look for alternative techniques in recent times \cite{lopesnetwork}.

\vspace{2mm}
\emph{Application to networking:} Header Space Analysis (HSA) \cite{kazemian2012header} is an example ternary symbolic simulation implementation proposed recently for verifying various properties such as reachability, loop detection, etc. for SDNs. Canini et al. have proposed a symbolic execution and model checking based  NICE framework for catching bugs which works by exploring symbolically all possible code paths \cite{canini2012nice}. In another work, Bishop et al. \cite{bishop2006engineering} have proposed symbolic evaluation testing of TCP implementation against a HOL specification.

\section{Programming Languages and Verification}
\label{sec:languages}

There are three ways to establish the meaning, or \emph{semantics}, of computer programs \cite{nielson2007semantics}. In \emph{operational semantics}, the program is modeled by execution on an abstract machine---this interpretation is useful for implementing compilers and interpreters. In \emph{axiomatic semantics}, pioneered by Hoare and Floyd, the program is modeled by the logical formulas it obeys---this interpretation is useful for proving program correctness. In \emph{denotational semantics}, the program is modeled by mathematical objects---this interpretation is useful for developing theoretical foundations of programming.


In the remainder of this section, we will discuss the grammar of languages, declarative programming, logic programming, and functional programming in sections \ref{sec:grammar}, \ref{sec:declarativeprogramming}, \ref{sec:logicprogramming}, and \ref{sec:functionalprogramming}, respectively.

\subsection{Grammar of Languages}
\label{sec:grammar}

The most common type of grammar used for specifying languages is known as the \emph{context-free grammar}. Context-free grammars are expressive enough to capture the recursive syntactic structure of most languages of our interest. The core component of a context-free grammar is a set of rules where a rule typically defines a name and an expansion for that name. The Backus-Naur form (BNF) is a formal notation used for encoding the grammar of a language in a form amenable to human consumption. The BNF notation is used by many programming languages, protocols or formats in their specification. A rule of the BNF notation has the following structure: ``$name ::= expansion$'' where the symbol $::=$ means `expands to' or `may be replaced with'. Every name in BNF is enclosed in angle brackets, $<>$. Choice is indicated by a vertical bar, $|$. For more details about the BNF format, the interested reader is referred to \cite{huth2004logic}. The BNF format is used to specify network programming languages in FlowExp (short for Flow Expression) \cite{kazemian2013real}, NetCoreLib for Frenetic \cite{foster2011frenetic} \cite{guha2013machine}, etc.

\subsection{Declarative Programming}
\label{sec:declarativeprogramming}

Declarative programming is a programming style in which we specify what the program must do without specifying how to do it. The imperative programming style adopted by imperative languages such as C, Java, etc., in contradistinction, focuses on specifying algorithmically how the computer must do its job. It may be highlighted that the imperative programming style harmonizes with the imperative procedural (how to) approach typically adopted in computer science while the declarative programming style dovetails with a mathematical or logic-based approach which emphasized declarative (what is) knowledge \cite{abelson1985structure}. Imperative programming style involves the use of mutable state variables which makes reasoning and verification a difficult task. Declarative programming style, in contrast, eschews maintenance of state variables and avoids invisible side-effects and relies instead of mathematical logic and evaluation of mathematical functions and logic formulae. Declarative programming is intimately tied to mathematical logic---programs in a declarative frameworks can be thought of as theories of formal logic, and computations as deductions in that logic space. Examples of declarative languages include SQL, frameworks such as: functional programming languages, logic programming languages, constraint logic programming, etc. In recent times, there has been a lot of interest in declarative languages, and in their use in networking especially cloud networking \cite{alvaro2010boom}, since declarative languages are well-suited to parallel programming\footnote{Almost every successful large-scale application of parallelism, e.g., SQL server, LINQ, MapReduce, etc., has been declarative and value-oriented \cite{jonesfuture}. This trend bodes well for the use of declarative programming, especially functional programming, in parallel computing.} \cite{hellerstein2010declarative}. Adoption of declarative programming languages is a manifestation of a contemporary trend in networking, brought on by the need to fix an ailing inflexible network architecture---and by software defined networking in particular, in which advanced programming techniques and database techniques are increasingly being applied to networking \cite{loo2009declarative}. We present two examples of SDN declarative languages: \emph{i)} the flow management language (FML) is a declarative language for SDN \cite{hinrichs2009practical} designed for network operators so that static network policies may be written and maintained more efficiently; \emph{ii)} the NetCore language \cite{guha2013machine} is a high-level declarative language proposed for SDN that allows programmer to describe what behavior is desired and not necessarily describe how to realize the implementation of that behavior. We will discuss SDN programming languages in more detail in section \ref{sec:sdnprogramminglang}.

\vspace{2mm}
\subsubsection{Logic Programming}
\label{sec:logicprogramming}

Logic programming provides many advantages including programmability at a very high level and natural support for formal semantics. Logic languages, such as Prolog, Lisp, have been very popular in the AI community for knowledge representation and automated reasoning. Prolog, an example declarative logic programming language designed primarily for AI based systems, works by stating and querying the logical relations between entities. Prolog like languages are also useful in formal verification for automated theorem proving. Logic programming languages are also popular in the databases community of computer science due to their support for declarative querying and symbolic manipulation. Datalog, an example database-based logic programming language, facilitates declarative definition of properties and relationships between objects with the language framework providing support for computing with these objects (including querying about objects declaratively).

The declarative programming style is superior to the procedural style in some significant ways---especially, in the context of formal verification \cite{maier1988computing}. The declarative style emphasizes the intent of a program and the static description of relationships and properties that hold in a program regardless of the computing context, thus easing understanding a computer program and reasoning about it. Unlike procedural languages, the effect of logic programming statements  is not dependent on the context (i.e., the state of the computer when the preceding statements were executed).

There has been a lot of interest in using declarative logic programming languages for simplifying the implementation of Internet protocols. They have been previously used for writing parsers (like the `yet another compiler-compiler' (yacc) parser tool \cite{johnson1975yacc}) for application layer protocols \cite{pang2006binpac}, declarative routing \cite{loo2005declarative}, and declarative networking \cite{loo2009declarative}. The basic insight behind declarative networking is the realization that recursive query languages are a natural fit for network protocols which essentially deal with computing and maintaining distributed state (such as information about routes, sessions, etc.) across the network. Network Datalog (NDlog) is a data and query model that has been proposed for declarative networking. NDlog, which implements a network specific subset of Datalog and supports distributed programming, exposes the partitioning of data across nodes and the link graph of the network. This makes the implementation much more amenable to static analysis and verifiable using other formal verification techniques such as general-purpose theorem provers. It has been shown that declarative implementations of popular protocols can be done much more concisely and efficiently while also allowing extensibility and safety \cite{loo2009declarative}. Logic programming languages have recently been proposed for SDNs \cite{katta2012logic}, FlowLog: \cite{nelson2013balance}, with researchers also exploring declarative network verification \cite{wang2009declarative}. In another work, Kazemian et al. have implemented a FML-like language in a Prolog frontend to enable network administrators to specify high-level policies \cite{kazemian2013real}.

\vspace{2mm}
\subsubsection{Functional Programming}
\label{sec:functionalprogramming}

The functional programming paradigm considers computation to be the evaluation of mathematical functions---that are not dependent on state and will always provide the same output for the same input. The main reason for the importance of functional programs is due to their direct correspondence with mathematical objects, which makes it easier to reason about them \cite{harrison1997introduction}. The functional paradigm avoids variables, or more technically---mutable state (i.e., variables whose values can be changed), and encourages a function-based programming worldview instead. By avoiding mutable state, the source of numerous subtle bugs in imperative-style programming languages, verification of programs become more simple. In functional programming, execution of a program means evaluation of the expression represented by the functional program. The functional programming style makes no use of variables. Instead of loops, the functional program makes use of recursive functions (i.e., functions that are defined in terms of themselves).

The main downfall of imperative programming is in race conditions when concurrency is supported.  Race conditions are much harder to detect and fix since they may arise of non-deterministic interleavings of concurrent threads (which may interleave in a myriad different ways). Imperative programming is always vulnerable to race conditions since it relies on mutable state.  Functional programming puts up much better with such race conditions since a pure functional language has no mutable state.  Since the future of programming is in concurrency and parallelism, functional programming is increasingly migrating from fringes of the programming world to the mainstream \cite{jonesfuture}. In summary, mutations allowed in the imperative programming paradigm severely limit any opportunities for automatic parallel execution, while the lack of dependencies in the (purely) functional paradigm presents great opportunities for automatic parallel execution.

The functional programming is based on the theoretical underpinnings of Alonzo Church's \emph{lambda calculus} \cite{scott2012lambda}, proposed in the 1930's, defines rules about using unnamed functions for representing and evaluating expressions. Lambda calculus, although originally intended as a formal logical system for mathematics is in fact a completely general programming language and defines a family of prototype programming languages. Many modern programming languages C++, Python, JavaScript, Ruby, Java 8, etc. borrow from the programming style of lambda calculus, following the lead of the Lisp programming language---which was the first mainstream language to include anonymous functions known as lambda functions. Two important features of lambda calculus is that it is functional---i.e., it is based on the concept of a mathematical function and include notation for function application and abstraction---and that is higher-order---i.e., it provides a systematic formalism and notation to deal with operators whose input and output may be other operators. The lambda calculus model significantly differs from the Turing model of a store with evolving state \cite{herken1995universal}. Interestingly, the Turing model and lambda calculus were invented in the same year, 1936. Turing showed in 1937 that both these models were equivalent and in fact defined the same class of computable functions. In any case, computer programs and mathematical proofs are directly related as the system of formal logic and computational calculi are analogous---the famous ``Curry-Howard correspondence'' expresses the isomorphism between proof structures and functional spaces \cite{luo1994computation}.

Popular functional programming languages include Lisp, invented by the AI pioneer John McCarthy, Haskell \cite{thompson1999haskell}, Caml, OCaml, Scala, etc. Historically, most successful languages have been written for specific purposes---e.g., Lisp was created for artificial intelligence, Fortran for numerical computation, and Prolog for natural language processing. The \emph{raison d'etre} for ML has been the need of an efficient language for theorem proving \cite{paulson1996ml}. ML originated as the metalanguage (thus its name ML) of the famous theorem proving system called Edinburgh LCF \cite{gordon1979edinburgh} for writing theorem proving algorithms in formal deductive calculus. ML was designed to have the full power of higher-order functional programming so that it could represent necessary inference rules and proof strategies. Since early time, functional languages and theorem proving (and formal verification in general) have been intimately intertwined. (Edinburgh) ML has spawned a wide range of ML-based descendant languages including Standard ML (SML) and OCaml. OCaml is a programming language specifically designed for writing theorem provers, with numerous major systems being written in it (e.g., SLAM verification system from Microsoft, HOL Light theorem prover, etc.). The OCaml language, being perfectly suited for symbolic manipulations, is used extensively by the Coq proof assistant which is used extensively for the verification of purely functional programs. Similarly, the SLAM verification system, proposed by Microsoft, also used OCaml programming language. The ACL2 (``A Computational Logic for Applicative Common Lisp'') theorem prover is also composed of a first-order, purely functional subset of Common Lisp.

With the recent paradigm shift in networking brought on by SDN, a clear trend of preferring high-level declarative languages, domain-specific languages (DSL), functional languages---and more specifically, \emph{functional reactive programming}---for programming SDNs (both the SDN controller as well as SDN applications) is emerging. Much of the recent work in SDN programming has followed the declarative programming coupled with the FRP paradigm  \cite{voellmy2012procera} \cite{foster2011frenetic} \cite{monsanto2013composing}. This trend is helped by ample foundational research in these fields in the programming and databases community, and by the verifiability properties of functional languages.

Nettle \cite{voellmy2011nettle} is a SDN specific language implemented as a domain-specific language in the functional programming language Haskell. Nettle adopts the design methodology of domain-specific languages (DSL) research, and is built in the paradigm of \emph{functional reactive programming} (FRP) \cite{wan2000functional}.  Nettle has been used for providing a comprehensive abstraction calculation constructs for configuring BGP policies. In a similar work, Procera \cite{voellmy2012procera} is a domain-specific language embedded in Haskell that can be used to specify  high-level dynamic reactive network control policies. In other work, the Frenetic project \cite{frenetic} defines a family of domain-specific languages for specifying high-level network policies. In the initial work in the Frenetic project \cite{foster2011frenetic}, two sub-languages were proposed: \emph{i)} a high-level declarative network query language---which enable Frenetic programs to read the network state using constructs for filtering, grouping, splitting, limiting, aggregating, etc., and \emph{ii)} a general purpose FRP-based network policy management library---using which the policy to govern the forwarding of packets through the network can be defined. The Frenetic framework borrows extensively from the FRP languages like Yampa \cite{hudak2003arrows}, etc., and reuses many of the proposed primitives. In more recent work, Pyretic \cite{monsanto2013composing} is an example DSL in the Frenetic family which supports composable policies constructed from a set of fundamental constructs such as basic policies and combinators along with associated techniques for compiling these techniques to OpenFlow switches.

\section{Applications of Formal Methods in Communication Networks}
\label{sec:applications}

In the history of the Internet, formal correctness has mostly taken a backseat to practical expediency and pragmatic considerations. The development, standardization, and deployment process is cumbersome and inflexible leading to an environment which only just works \cite{handley2006internet}. As an example of the unfortunate adhocism that pervades the culture of network protocols, it is noted that BGP, despite any lack of convergence guarantees, is often used in service as an interior-gateway routing protocol (IGP) \cite{griffin2005metarouting}. While there were some initial successes in the application of formal methods to networking \cite{zave2012formal}, the networking enterprise quickly transformed into a complex behemoth impervious to any attempts at formal analysis and verification. In addition to the inherent complexity of networking protocols, the vertical integration of control and data planes meant lack of modularity and a paucity of useful abstractions \cite{rexford2012report}. With the tools of formal methods unable to tame the staggering complexity of networking, the resulting frustration bred skepticism leading to a widespread critical view, enunciated by Vint Cerf \cite{vintcerf}, that formal methods are ``overblown, verbose, hard to use, (and) hard to understand''. Fortunately, modern attempts at redesigning the Internet ossified architecture---and more specifically, the SDN movement---create new abstractions by separating the control and data planes and thus allow a great opportunity for incorporating formal methods in networking. With the utility of Internet firmly entrenched in all aspects of modern life, the use of formal methods for ensuring correctness of specification and operation is anticipated to be incorporated into mainstream Internet operations.

In the remainder of this section, we will describe various applications of formal verification methods to networking. We will discuss protocol verification in section \ref{sec:protocolverification}. We will follow it up in section \ref{sec:propertyverification} with a discussion on property verification and discuss reachability analysis, loop detection, and isolation verification, in sections \ref{sec:reachabilityanalysis}, \ref{sec:loopdetection}, and \ref{sec:isolationverification}, respectively. We will then discuss the use of formal verification for network configuration management, and network security in sections \ref{sec:netconfmanagement} and \ref{sec:netsecurity}.  We will discuss various issues related to generic network verification in section \ref{sec:networkverification}; in particular, we will discuss declarative verification, hardware verification, formal specification and synthesis, and implementation variation in sections \ref{sec:declarativenetverification}, \ref{sec:hardwareverification}, \ref{sec:formalspecificationandsyn}, \ref{sec:implementationverification}. Finally, we will wrap up this section with application of formal methods specifically to SDN in section \ref{sec:sdnapplications}. More specifically, we will discuss the new opportunities created by SDN in section \ref{sec:whatisnewaboutSDN}, and follow it up with discussions on SDN programming languages, data plane verification, control plane verification, and network debugging in sections \ref{sec:sdnprogramminglang}, \ref{sec:dataplaneverification}, \ref{sec:controlplaneverification}, and \ref{sec:networkdebugging}, respectively. A tabulated summary of various applications of formal methods in the context of networking is presented in table \ref{tab:networking}. A summary of various tools that are used in this regard is presented separately in table \ref{tab:tools}.

\begin{table*}
\caption{Summary of \textbf{applications} of formal verification in networking}
\centering
\begin{tabular}{p{3.3cm} p{3.3cm} p{10.3cm}}
\toprule
\textbf{\emph{Project and Reference}} & \textbf{\emph{Technique}} & \textbf{\emph{Brief Summary}} \\
\midrule\\
\multicolumn{3}{l}{\textbf{\emph{Protocol Verification}}} \\
Bishop et al. \cite{bishop2006engineering} & Symbolic Evaluation & Proposed symbolic evaluation testing of TCP implementation against a HOL specification\\
Ridge et al. \cite{ridge2008rigorous} & HOL proof assistant & Proposed a rigorous approach for modeling and verifying TCP using the HOL proof assistant\\

\hline\\
\multicolumn{3}{l}{\textbf{\emph{Reachability Analysis}}} \\
Xie et al.\cite{xie2005static} & Static Checking & Proposed a graph-theoretic algorithm (transitive closure) for static analysis of IP networks with support for ACL policies\\
Khakpour et al. \cite{khakpour2010quantifying} & Static Checking & Proposed a tool Quarnet comprising algorithms for quantifying reachability based on network configuration (incorporated ACL) and for querying network reachability\\
Al-Shaer et al. \cite{al2009network} & (Symbolic) Model Checking & Proposed a BDD/ CTL based symbolic model checking approach for performing `network configuration in a box'\\
Lopes et al. \cite{lopesnetwork} & SAT solvers & New SAT based solutions for the reachability set predicate\\
HSA \cite{kazemian2012header}  & SAT solvers; Static Checking & HSA provides a protocol-agnostic method for finding data plane bugs in networks by jointly studying the header space of packets and transformations applied to it by networking boxes \\
NetPlumber \cite{kazemian2013real} & SAT solvers; Static Checking & HSA-based real-time policy checker for networks that works with incremental recomputation \\
VeriFlow \cite{khurshid2012veriflow} & Mininet, Depth-first search, Tries &  Implemented as a layer between SDN controller and switches, VeriFlow verifies network-wide invariants in real-time dynamically  as a forwarding rule is added\\
AP Verifier \cite{yangreal} & Atomic Predicates Verifier & AP verifier reduces the set of predicates representing packet filters to minimal atomic predicates, using which AP verifier dramatically improves computation of network reachability\\

\hline\\
\multicolumn{3}{l}{\textbf{\emph{Loop Detection}}} \\
HSA, NetPlumber, VeriFlow, AP Verifier  & See above & These tools provide support for loop checking as well. \\

\hline\\
\multicolumn{3}{l}{\textbf{\emph{Isolation Verification}}} \\
AP Verifier  \cite{yangreal} & Atomic Predicates Verifier & AP verifier, discussed above, can also be used to verify slice isolation \cite{yangreal}\\
``Splendid Isolation'' \cite{gutz2012splendid} & Model Checking & Proposes a slices abstraction for SDNs with automatically verifiable formal isolation properties (expressed in CTL and checked through NuSMV tool)\\


\hline\\
\multicolumn{3}{l}{\textbf{\emph{Configuration Management}}} \\
rcc \cite{feamster2005detecting} & Static Analysis & Static analysis tool for detecting BGP configuration faults proactively before deployment \\
Qie et al. \cite{qie2004using} & Static Analysis & Proposed using service grammar, incorporating a requirements language containing global high-level constraints, for detecting BGP configuration errors \\
Narain et al. \cite{narain2005network} & (Lightweight) Model Checking; Scenario Finding & Proposed a method for managing (i.e., formalizing, and automating, reasoning about) network configuration with `model finding' using the Alloy analyzer \\
ConfigChecker  \cite{al2009network} & (Symbolic) Model Checking & Performs firewall verification with BDD-based model checking to perform symbolic reachability analysis \\
FlowChecker \cite{al2010flowchecker} & (Symbolic) Model Checking & The FlowChecker tool can be used to verify the correctness of OpenFlow federated infrastructures and debug reachability and security problems\\
Anteater \cite{mai2011debugging} & SAT solvers; Static Checking & Anteater, builds upon Xie et al. \cite{xie2005static}, \emph{implements} a tool for checking invariants in the data plane by transforming invariants into SAT instances to be checked against network state by a SAT solver \\
Paulson et al. \cite{paulson2002isabelle} & Theorem Prover & BGP policy verification was performed using Isabelle/ HOL prover\\

\hline\\
\multicolumn{3}{l}{\textbf{\emph{Network Security}}} \\
FLOVER \cite{sonmodel} & Model-Checking; SMT solvers &  Verifies that the aggregate of flow policies instantiated within an OpenFlow network does not violate the network’s security policy \\
MulVAL \cite{ou2005mulval} & Logic-based Analysis & A logic-based network security analyzer \\
Zhang et al. \cite{zhang2012icnp} & SAT and QBF solvers & A SAT based technique for comparing the equivalence and inclusion relationship between two firewalls, and also propose Quantified Boolean Formula (QBF) based ACL optimization\\
Margrave \cite{nelson2010margrave} & SAT solvers \& Scenario Finding & Firewall analysis tool that allows tracing behavior to specific rules and verification against security goals\\
Al-Shaer et al. \cite{al2004discovery} & Tree based model & Proposed a ``Firewall Policy Advisor'' for managing firewall filtering rules, and for detecting all anomalies in single or multiple firewall environments \\
Kothari et al. \cite{kothari2011finding} & Symbolic Execution \& Model Checking & Studies protocol manipulation attacks in which adversaries induce honest players into undesirable behaviors\\
Ritchey et al. \cite{ritchey2000using} & Model Checking & Uses model checking to analyze network vulnerabilities \\
Gouda et al. \cite{gouda2007structured}& Firewall Decision Diagrams & Presented a structured firewall design ensuring consistency, completeness and compactness. Also, proposed firewall decision diagram (FDD) for modeling firewall specification formally\\

\hline\\
\multicolumn{3}{l}{\textbf{\emph{Automatic Synthesis}}} \\
FVN project \cite{wang2009formally} & Logic-based framework & FVN presents a approach towards unifying the design, specification, implementation, and verification of networking protocols based on a logic language NDLog \\
Noyes et al. \cite{noyestoward} & Model Checking & Proposed techniques for synthesis of network updates using NuSMV and OCaml tools \\
Wang et al. \cite{wangautomated2013} & Reactive Synthesis \& Model Checking & Proposed techniques for automated synthesis of reactive controllers for SDNs\\

\hline\\
\multicolumn{3}{l}{\textbf{\emph{Data Plane Verification}}} \\
FlowChecker \cite{al2010flowchecker}, Anteater \cite{mai2011debugging}, HSA \cite{kazemian2012header} & Static Checking & These tools perform static verification of the data plane of a network based on a snapshot of network state\\
NetPlumber \& VeriFlow \cite{khurshid2012veriflow} & Dynamic Checking & These tools perform dynamic verification of the data plane using incremental recomputation techniques \\
ATPG \cite{zeng2012automatic} & Automatic Testing & ATPG is an automatic testing tool for generating test packets\\
NetSigtht, ndb \cite{handigol2012debugger} & Interactive Debuggers & Interactive debugging tools that operates passively\\

\hline\\
\multicolumn{3}{l}{\textbf{\emph{Control Plane Verification}}} \\
Scott et al. \cite{scott2013did} & `Retrospective Causal Inference' & Proposed improving SDN troubleshooting by automatically identifying the minimum sequence of inputs responsible for causing a control software bug\\
Guha et al. \cite{guha2013machine} & Theorem Proving &  Proposed a featherweight version of the OpenFlow protocol, and used the Coq tool for verifying the network controller \cite{guhaformal}\\
Reitblatt et al.  & Theorem Proving & Ensuring per-packet and per-flow consistency of network updates using the Coq prover \\
NICE \cite{canini2012nice} & Symbolic Execution \& Model Checking & Performs symbolic execution of OpenFlow applications while applying model checking to explore the entire state space of the network \\
FlowLog \cite{nelson2013balance} & Model Checking & Proposed a declarative finite-state language for programming SDN controllers that balances expressiveness and analysis and is amenable to model checking\\
Sethi et al. \cite{sethiabstractions} & Model Checking & Proposed new abstractions for model checking SDN controllers
\\ \bottomrule
\end{tabular}
\label{tab:networking}
\end{table*}

\begin{table*}
\caption{Representative summary of various formal verification \textbf{tools}}
\centering
\begin{tabular}{p{2.2cm} p{2.2cm} p{12.5cm}}
\toprule
\textbf{\emph{Technique}} & \textbf{\emph{Tool}} & \textbf{\emph{Brief Summary}} \\
\midrule
\multicolumn{3}{l}{\textbf{\emph{Model Checkers}}} \\
& SPIN \cite{holzmann1997model} & SPIN is a mostly automated tool for verifying distributed and concurrent systems\\
& NuSMV \cite{cimatti2002nusmv} & NuSMV, an extension of the original symbolic model checking tool SMV, is a model checking tool based on BDDs \\
& Alloy \cite{jackson2006software} & Alloy analyzer is a light-weight formal method that can analyze user specified properties of a (partial) model\\
& PRISM \cite{kwiatkowska2009prism} & PRISM is a probabilistic model checker suitable for systems that exhibit probabilistic behavior\\
& UPPAAL \cite{larsen1997uppaal} & UPPAAL is a model-checking based tool-box, based on timed-automata formalism, used for verification of real-time systems\\
\multicolumn{3}{l}{\textbf{\emph{Theorem Provers}}} \\
& Edinburgh LCF \cite{gordon1979edinburgh} & Interactive theorem prover proposed in 1972 which introduced ML language as a metalanguage for writing proving tactics\\
& HOL \cite{gordon1987hol}& HOL represents a family of interactive theorem provers that are based on higher-order logics and strategies\\
& Isabelle \cite{paulson2002isabelle}& A popular LCF-style theorem prover (written in Standard ML) that can work with various logics\\
& ACL2 \cite{brock1996acl2}& ACL2 is a mechanical theorem prover with a Common Lisp-variant programming language, and an extensive first-order logic based theory\\
& PVS \cite{owre1992pvs}& PVS is an automated theorem prover with an integrated specification language with multiple support tools\\
& Coq \cite{bertot2004interactive}& Coq is an interactive theorem prover that assists in finding proofs, and in extracting a certified program from the constructive proof\\
\multicolumn{3}{l}{\textbf{\emph{SAT and SMT solvers}}} \\
& Microsoft's Z3 \cite{de2008z3}& Z3 is a state of the art SMT solver from Microsoft Research\\
& Kodkod \cite{torlak2007kodkod} &  Kodkod is a SAT-based constraint solver that can work with first-order logic with relations, transitive closure, etc. \\
& YICES \cite{dutertre2006yices}& Yices is an efficient SMT solver that can also act as a SAT and MaxSAT solver\\
\\ \bottomrule
\end{tabular}
\label{tab:tools}
\end{table*}

\subsection{Protocol Verification}
\label{sec:protocolverification}

In layered communication networks, protocols define the set of rules governing exchange of messages between interacting processes which serve two related goals: firstly, to provide service to the local protocol layers above, secondly, to interact according to a defined protocol with remote peer partners on other machines. In terms of specification, the former goal is defined through service specification, while the latter is defined through protocol specification. Both these specifications---service-specification and protocol-specification---can be verified against their design or implementation. Verification at the design stage is more useful as it can avoid unnecessary incorrect implementation \cite{yuang1988survey}.

Holzmann  \cite{holzmann1991design} \cite{holzmann1993design} lists three ways in which protocols can fail: \emph{deadlocks}---when all the protocols stall waiting for conditions that can never be fulfilled; \emph{livelocks}---when execution sequences keep getting repeated indefinitely without the protocol making any effective progress, and \emph{improper terminations}---when the protocol completes execution without satisfying the proper terminating conditions. The general problem of finding deadlocks in protocols is known to be complex, i.e., \PSPACE-complete at best which makes it undecidable for unbounded message queues. Thus any method that relies exclusively on an exhaustive search of state space method is bound to fail, thus prompting much research on alternate non-exhaustive methods that exploit symmetry and abstraction. Also, due to the inherent complexity of the problem, we set a more conservative target in protocol verification of detecting the presence of errors---should they exist---with high probability instead of striving to prove the absence of errors with certainty.

Various works have been proposed for protocol verification including rigorous specification and conformance testing techniques for network protocols \cite{bishop2005rigorous}, rigorous treatment of the TCP protocol \cite{bishop2006engineering} \cite{ridge2008rigorous}, verification of ad-hoc routing protocols for wireless sensor networks: \cite{chen2013review}. There have been a few survey papers written focusing on communication protocols \cite{lai2002survey} \cite{bochmann1980formal},  including a FSM-based protocol verification survey \cite{yuang1988survey} and a survey documenting experience with protocol description \cite{zave2011experiences}. Various tools have been used for protocol verification including the theorem proving tool Isabelle \cite{paulson2002isabelle} for BGP policy verification and the proof assistant Coq tool \cite{bertot2004interactive} for creating a featherweight version of the OpenFlow protocol \cite{guhaformal}.

\subsection{Property Verification}
\label{sec:propertyverification}

There is great interest and intent in the research community to develop technological support for automatic verification of various properties of protocols and systems. When we are verifying the property of a system, we are essentially interested in two kinds of properties: \emph{i) Safety property} where we are mainly interested that `bad' things will not occur, \emph{ii: Liveness property:} where we are mainly interested in that `good' things will eventually occur \cite{camurati1988formal}. In general, safety property bugs are easier to discover by finding a counterexample, while liveness property violations are difficult to obtain---in particular, a liveness violation example would require finding an infinitely long execution trace in which the desired `good' property never happens. \cite{killian2007life}. Recent works such as Anteater \cite{mai2011debugging}, Header Space Analysis (HSA) \cite{kazemian2013real}, FlowChecker \cite{al2010flowchecker}, VeriFlow \cite{khurshid2012veriflow} use an automatic solver to check properties of a logical representation of switch configurations.

In the remainder of this section, we will cover example properties of reachability analysis, loop detection, and isolation verification, packet destination control.

\vspace{2mm}
\subsubsection{Reachability Analysis}
\label{sec:reachabilityanalysis}

Reachability analysis is a powerful method widely used for formal verification of protocols \cite{holzmann1988improved} and concurrent distributed systems.  Unfortunately, reachability analysis suffers, like all methods based on finite state machines, from the state-explosion problems. Reachability analysis can benefit from symbolic methods which work without inspecting all the reachable states of the system to scale to large networks---e.g., BDD-based symbolic traversals have been proposed for reachability analysis of large finite state machines \cite{cabodi1997improved}. An example work that utilizes BDD-based symbolic model checking for reachability analysis is the `Network configuration in a box' project by Al-Shaer et al. \cite{al2009network}.

In a networking context, reachability analysis was first proposed for IP networks by Xie et al.\cite{xie2005static}. The technique proposed utilized a static snapshot of network configuration,  culled from configuration state from each of the network routers, for determining reachability between applications running on end-hosts. This reachability information is very useful in network troubleshooting and management for verifying the implementation of the intent of the network designer, and for troubleshooting reachability problems. Xie et al. reduce the reachability problem to a classical graph theoretic problem which can be solved in polynomial time by computing the transitive closure\footnote{Transitive closure of $G$ has an directed edge from $x$ to $y$ iff there is a directed path from $x$ to $y$ in $G$. Transitive closure is a standard graph-theoretic technique which, intuitive speaking, provides an efficient method for answering the reachability question `where can we get from here?'} to set union and intersection operations on the representation of reachability set. Recently, advances in SAT technology has led to its use for reachability analysis problems \cite{zhang2013sat} \cite{lopesnetwork}.

Kazemian et al. have proposed a general protocol-agnostic static checking framework for networks based on header space analysis (HSA) \cite{kazemian2012header}. Kazemian et al. have proposed a library of tools, called Hassel, which implements their proposed HSA based framework to identify important classes of failures which also includes forwarding loop detection, traffic isolation failure, beside reachability failure. The basic insight of HSA is to model a packet by its header by treating the entire header field as a concatenation of bits without any associated semantics---instead, the packet may be considered as a (geometric) point in the ${0,1}^L$ geometric space where $L$ is the maximum length of the packet header. The network is then modeled as being composed of network boxes (such as routers, switches, etc.) that transform packets from one point to another point, or possibly set of points (assuming multicast). This geometric approach taken by HSA (i.e., of representing the packet as a point in a subspace) allows the Hassel tools to work in a protocol-agnostic manner. Using HSA, we can easily \emph{i)} find all packets that can reach from a point A to another point B, \emph{ii)} find loops regardless of the protocol/ layer, \emph{iii)} prove that two slides are isolated. Unlike model checking, HSA is not limited to providing a single counterexample in case of a failure detection, but can importantly provide information about the full set of failed packets.

Lopes et al. \cite{lopesnetwork} have recently proposed extending the reachability predicate (``Can a packet from node A reach node B?'') to a generalized abstraction of \emph{reachability set} (``What are all the packets that can reach node B from node A?''). It is highlighted in \cite{lopesnetwork} that reachability sets are useful for two reasons: \emph{incremental computation} and \emph{intelligibility}. In general, the tools for calculating reachability sets are less developed, although some languages like Datalog provide out of the box support for computation of reachability sets. The technique of incremental computation is useful for dynamic verification (i.e., when a new rule is being added) and has been recently proposed for real-time verification of SDN networks \cite{khurshid2012veriflow} \cite{kazemian2013real}. The main insight underlying such an approach is the realization that a single rule change is unlikely to change the underlying network state machine drastically. Therefore, small modifications are necessary to the ``reachability set'' to incorporate the changes introduced by the addition of the new rule. Reachability set is also more intelligible as it produces a more general counter example---e.g., it can provide a set of packets being dropped.

In a promising recently proposed work \cite{yangreal}, Yang and Lam present ``Atomic Predicate (AP) Verifier'', which reduces the set of predicates representing network packet filters to a set of atomic predicates that is provably both minimum and unique, which can be used to dramatically improve the computation of network reachability. The basic insight of this work is that atomic predicates have the following key property: Any given predicate is equal to the disjunction of the a subset of atomic predicates, and thus can be stored and represented as a set of integers identifying the atomic predicate. The conjunction (or the disjunction) of two predicates can be computed quickly as the intersection (or union) of two sets of integers. As an example, Yang and Lam show that while the Stanford network has 71 ACLs and 1584 rules, there were only 21 atomic predicates for these ACLs and rules (due to great redundancy in the forwarding and ACL rules). By encoding the rules in terms of atomic predicates (in the form of BDDs which can be manipulated through well-known graph BDD algorithms), this unnecessary redundancy is removed leading to much greater space and time efficiency. In their performance evaluation, Yang and Lam compare AP Verifier with Hassel in C and NetPlumber to demonstrate that AP verifier is significantly more time and space efficient \cite{yangreal}.

Property verification also includes questions about \emph{packet destination control:} Can a packet \emph{i)} get out of the network, \emph{ii)} get dropped, \emph{iii)} go through certain switches, or \emph{iv)} never pass through certain links. Model checking as well as ternary symbolic simulation techniques can be used for packet destination control \cite{zhang2012verification}.

\vspace{2mm}
\subsubsection{Loop Detection}
\label{sec:loopdetection}

Header Space Analysis (HSA) \cite{kazemian2012header} defines a ``network algebra'' which captures the manipulation of packet headers by network routers and switches. In the HSA framework, packet headers, represented as $n$-dimensional bit fields, are operated upon by the function defined by routers and switches which effectively transform the packet headers. HSA \cite{kazemian2012header}, and its enhancement NetPlumber \cite{kazemian2013real}, can verify a range of properties such as connectivity, reachability between ports, absence of any loops, and isolation between groups, etc.
Various other approaches have been proposed in literature for loop detection including ConfigChecker \cite{al2009network}, AP Verifier \cite{yangreal}, etc.

\vspace{2mm}
\subsubsection{Isolation Verification}
\label{sec:isolationverification}

For various reasons (such as security, confidentiality, etc.), it is sometimes desirable to ensure that certain kinds of traffic are isolated from each other. In current Internet, this is managed by various ad-hoc mechanisms often requiring manual intervention. For example, techniques used for ensuring isolation include: \emph{i)} low level mechanisms such as VLANs or ACLs requiring configuration, \emph{ii)} special purpose devices such as firewalls, \emph{iii)} or complex hypervisors such as the FlowVisor system \cite{sherwood2010carving} for OpenFlow networks. It is desirable to have more fundamental abstractions that can be exploited to provide verifiable isolation between traffic as desired. An initial work in this regard has been presented for SDNs in the ``splendid isolation'' project \cite{gutz2012splendid} proposed as part of the Frenetic project \cite{frenetic}. In this work, a slice abstraction is presented and algorithms for compiling slices is presented along with a tool for automatic verification of formal isolation properties. In other works, AP Verifier can also verify slice isolation as reported in \cite{yangreal}.

\subsection{Network Configuration Management}
\label{sec:netconfmanagement}

Configuration errors can create numerous connectivity, security, performance, and reliability problems. It has been pointed out in literature that the bulk of network downtime is in fact due to  manual errors \cite{networkdowntime} and misconfiguration of devices \cite{zhang2012verification}. The problem is especially acute since it is not far fetched for a misconfiguration of a single device to cripple an entire network. Various problems can arise from bugs due to misconfiguration including access control failures, isolation guarantee failures, routing loops, reachability failures, blackholes, etc. The presence of such problems can have debilitating effect on network performance and efficiency, thus motivating a more rigorous and formal management of network configuration. In configuration management, we would like to have multiple abstractions, incorporating correctness checks, between the high-level global end-to-end requirements and low-level distributed configuration at individual devices.

Static analysis has been used extensively for detecting configuration faults. Feamster et al. proposed a static analysis tool \emph{rcc} for detecting BGP configuration faults \cite{feamster2005detecting}. The \emph{rcc} tool allowed proactive analysis of network configurations before deployment in an operational network by checking that BGP configuration satisfies a set of constraints, based on the correctness specification. The \emph{rcc} tool, like most practical static analysis tools, is neither complete nor sound---i.e., it can miss problematic configurations, and may complain about harmless deviations from the best practices. Nevertheless, \emph{rcc} was able to find many important classes of errors to make it useful in practice. Qie et al. \cite{qie2004using} proposed an approach based on ``service grammar'' for BGP which incorporated a requirements language using which the network operator can specify high-level requirements  against which the system may be checked. Unfortunately, the proposed grammar was rather low-level thus having possibilities of erroneous specification. In another work, Narain et al. proposed managing network configuration through model finding \cite{narain2005network} while using the Alloy analyzer \cite{jackson2006software}. In yet other work, ConfigChecker  \cite{al2009network} performs firewall verification with BDD-based model checking to perform symbolic reachability analysis. Configuration management has been a fertile area for application for formal verification methods with various proposals in literature \cite{feamster2004practical} \cite{narain2005network} \cite{al2010flowchecker} \cite{mai2011debugging}.

\subsection{Network Security}
\label{sec:netsecurity}

There are various important subproblems of firewall verification and synthesis \cite{zhang2012icnp}. Firstly, the \emph{firewall equivalence checking} problem focuses on determining if two firewalls have identical behavior---i.e., they drop and permit the same set of packets. Secondly, the \emph{firewall inclusion checking} problem compares two firewall policies and can verify that one policy is inclusive, i.e., more strict, than the other policy. Thirdly, the \emph{firewall rule redundancy checking} problem focuses on determining redundant rules---i.e., rules that can be deleted without affecting the behavior of the firewall. Lastly, the \emph{firewall synthesis} problem focuses on synthesizing a firewall with minimum number of rules install that matches exactly the behavior of another given firewall.

There has been a lot of work in firewall verification and synthesis (e.g., \cite{zhang2012icnp}  \cite{nelson2010margrave} \cite{al2004discovery} \cite{kothari2011finding} \cite{gouda2007structured}) and vulnerability analysis (e.g., \cite{kothari2011finding} \cite{ritchey2000using} \cite{ramakrishnan2002model}) and a variety of techniques have been utilized including static analysis \cite{chess2007secure}, model based analysis \cite{ramakrishnan2002model} \cite{nicol2004model}, logic-based analysis \cite{ou2005mulval}, SAT solvers, model checking  \cite{sonmodel} \cite{ritchey2000using}, new abstractions (e.g., firewall decision diagrams or FDD \cite{gouda2007structured}, atomic predicates (AP) verifier \cite{yangreal}, etc.). The interested reader is referred to a detailed description of related work in \cite{nelson2010margrave} and \cite{gouda2007structured}.



\subsection{Network Verification}
\label{sec:networkverification}

Traditionally, the focus of formal verification community has been on hardware systems or software systems, and relatively less on network verification. Networked systems comprise a software component (implementing the node OS, protocols, applications, etc.) and a hardware component (featuring the range of hardware configuration such as microprocessors, general purpose processors, DSPs, ASICs, etc.). Networked system are in fact distributed systems composed of end hosts that use the network as well as networking nodes (such as routers, switches, and various middleboxes such as firewall, load balancers etc.) that implement the network. In previous work, network verification is considered as essentially a state machine verification problem \cite{lopesnetwork}---i.e., a communication network can be visualized as a finite network of FSMs. Although, this problem is quite complex theoretically---\PSPACE-complete for the general problem of verification of network of FSMs---structural properties of networks fortunately enable techniques like Anteater \cite{mai2011debugging} and HSA \cite{kazemian2012header} to work satisfactorily in practice.

\vspace{2mm}
\subsubsection{Declarative Network Verification}
\label{sec:declarativenetverification}

As pointed out earlier, there is an increasing trend in using declarative programming techniques, and techniques that have been successful in deductive databases community, in networking. The use of such techniques also enables importantly the ability to perform network verification. There has been some work in this regard \cite{wang2009declarative} in which the task of formal specification is performed through declarative networking code, using Network Datalog (NDLog), a distributed variant of Datalog, while verification is done through a general-purpose theorem prover.

\vspace{2mm}
\subsubsection{Hardware Verification}
\label{sec:hardwareverification}

Formal verification methods have been used for hardware verification of networking devices. A general survey of formal verification in hardware design can be seen at \cite{kern1999formal}. Some sample works in hardware verification in networking include verification of: \emph{i} the lookup machine of a hardware router \cite{antovs2004hardware}, \emph{ii} the Fairisle ATM swithing element \cite{curzon1994formal}, \emph{iii)} network-on-chip \cite{borrione2009formal} \cite{van2009towards}.


\vspace{2mm}
\subsubsection{Formal Specification and Synthesis}
\label{sec:formalspecificationandsyn}

There are many benefits in formally specification including the clarity accompanying  rigorous specification of high-level specification of the target networking problem along with the ability to employ mechanized correctness checking to weed out trivial mistakes through techniques such type-checking. It has been shown in research that informal specification of protocols can lead to incorrect reasoning and implementation \cite{zave2012using} and ambiguity \cite{zave2008understanding}.

In order to create a correctly performing implementation, it is worthwhile to invest time and effort in \emph{design verification}.  Various approaches can be explored including specialized meta-theories specific to routing and forwarding \cite{griffin2005metarouting}, axiomatic logic-based formalisms \cite{karsten2007axiomatic}, or declarative programming frameworks \cite{loo2009declarative} \cite{voellmy2011nettle}, to specify the design. These formalisms can then be analyzed used methods like theorem provers, model checking, SAT/ SMT solvers, lightweight formal methods etc. to verify the correctness of the design and thereby guide the implementation.


There also has been work in synthesizing protocol implementations from formal specifications. An example work in this regard is the ``formally verifiable networking'' (FVN) project \cite{wang2009formally}. In another work, the synthesis of network updates have been proposed \cite{noyestoward}. Recently Lopes et al. \cite{lopesnetwork} have indicated building a synthesis tool for Microsoft Azure firewalls as their future work---such a tool can enable synthesis of low-level rules from a high-level specification and thus network operators can forego the error-prone access control list (ACL) configuration CLI.

\vspace{2mm}
\subsubsection{Implementation Verification}
\label{sec:implementationverification}

Having studied techniques that can be used to verify design in previous subsections, we will now see that a variety of techniques, described earlier in section \ref{sec:formalverification}, can be used to verify \emph{implementations}. In particular, we can make use of static checking as well as dynamic checking. In static validation, correctness properties are defined as invariants or constraints which are then checked to find out any system faults. In certain cases, a pre-processing stage may be necessary to transform the real system into an intermediate more checkable form. Static analysis and model checking are static validation tools. While most model checking tools work with specification models, some model checking tools (such as MaceMC \cite{killian2007life}, VeriSoft \cite{godefroid1997model} and CMC \cite{musuvathi2002cmc}) can work directly with implementation code making them very valuable for verifying implementations. In dynamic validation, on the other hand, we rely on runtime verification and testing---which per se are not really formal verification tools but nonetheless perform a complementary role.

\subsection{Applications in SDN}
\label{sec:sdnapplications}

In this section, we will discuss new opportunities offered for incorporating programming and verification advances into the networking context by the SDN architecture. We will initially discuss the new degrees of freedom offered by SDN in section \ref{sec:whatisnewaboutSDN}. We will discuss SDN programming languages in section \ref{sec:sdnprogramminglang} and will thereafter talk about data plane and control plane verification in sections \ref{sec:dataplaneverification} and \ref{sec:dataplaneverification}, respectively. Finally, we will discuss SDN debugging tools in section \ref{sec:networkdebugging}.

\vspace{2mm}
\subsubsection{What is new about SDN?}
\label{sec:whatisnewaboutSDN}

In traditional networking, the complex intricacies of a vertically integrated network architecture largely ruled out applications of formal methods to the domain of networking. This resulted in ad-hoc management of networks by ``masters of complexity'' \cite{shenkerStanford}---network administrators who kept networks running mainly through intuition and judgment honed through experience with a very  limited tool-set. Fortunately, the recent SDN architecture is much cleaner and offers an opportunity at rethinking networking management and troubleshooting \cite{heller2013leveraging}. There are three reasons for the optimistic evaluation of verification prospects of SDNs: firstly, the control plane that previously ran as distributed algorithms across individual devices has now been refactored into a single program that runs on the controller; secondly, the heterogeneity in traditional networking---in devices, configuration interfaces, vendors, and softwares---has given way to stock programmable switches supporting standard interfaces with precise semantics \cite{guha2013machine}; lastly, it is envisioned that the core network, or the \emph{fabric}, in the new SDN architecture will be purely hardware (finite state) and is thus amenable to efficient application of verification techniques \cite{zhang2012verification}. These new degrees of freedom enabled by SDN have ignited a renewed resolve in the networking community of applying formal methods to networking and to put networking on a solid theoretical foundation \cite{guhaformal} \cite{shinformal} \cite{skowyra2013verifiably}.


\vspace{2mm}
\subsubsection{SDN Programming Languages}
\label{sec:sdnprogramminglang}

Various SDN specific programming languages have been proposed recently (e.g., Frenetic \cite{frenetic}, NetCore \cite{monsanto2012compiler} \cite{guha2013machine}, Pyretic \cite{monsanto2013composing}, and NetKat \cite{anderson2013netkat}, etc.). These network programming languages enable programmers, in line with the vision of software defined networks, to define the desired network behavior at a high-level and the compiler then translates the high level abstract description to rules that are installed on the underlying hardware devices. The NetCore language \cite{monsanto2012compiler} was initially designed to provide support for parallel composition and was later extended by Pyretic \cite{monsanto2013composing} for sequential composition. NetCore provides a rich set of programming primitives including predicates for filtering packets, actions for modifying and forwarding packets, and (parallel and sequential) composition operators for building elaborate policies from simpler ones. NetCore has even been formalized in Coq. NetKat is similar to NetCore and Pyretic, but additionally provides formal axiomatic semantics and a compiler based on an equational theory for reasoning about programs. NetKat is based on \emph{Kleene algebra with tests} which is a mature framework that combines Kleene algebra---useful for reasoning about network structure---and Boolean algebra which is useful for reasoning about the predicates that define switch behavior. NetKat provides consistent reasoning principles that other network programming languages lack. In contrast to fore-mentioned languages, which have a functional bent and are suited for programming of centralized controllers, the DataLog\footnote{Datalog is a declarative logic programming language used as a query language for deductive databases. It is a simplified form of Prolog, and can be envisioned as a subset of Prolog sans the complex terms allowed by Prolog.} based declarative network programming language \emph{NDLog} \cite{loo2005declarative} \cite{loo2009declarative} is a logic programming language suited to distributed programming.

\vspace{2mm}
\subsubsection{Data Plane Verification}
\label{sec:dataplaneverification}

Various approaches have been proposed for data plane verification including \emph{i)} static checking--in which the correctness is verified independently, \emph{ii)} dynamic checking---in which new forwarding state is checked before being added, \emph{iii)} automatic testing---where the correct behavior of the dataplane is checked automatically, and \emph{iv)} interactive debugging---which aims at finding bugs in operational networks. The Anteater \cite{mai2011debugging}, FlowChecker \cite{al2010flowchecker}, and Hassell \cite{kazemian2012header} tools are example \emph{static checking tools}. Various real-time \emph{dynamic checking tools} have been proposed in literature including NetPlumber \cite{kazemian2013real} and VeriFlow \cite{khurshid2012veriflow}. The NetPlumber tool uses a novel header space analysis for performing a real time network policy check, while the VeriFlow tool verifies network invariants---e.g., lack of access control violations, absence of routing loops, blackholes, etc.---in real time and presents a diagnostic report in case of a violation. The Automatic Test Packet Generation (ATPG) tool is an automatic testing tool that automatically generates test packets \cite{zeng2012automatic}. The ATPG verifies full reachability in a network, using minimal network of test packets by using a heuristic solver for the min-set-cover problem, and detects anomalies by looking for persistent packet drops that are indicative of some software or hardware errors.
Finally, the NetSigtht and the Network Debugger (ndb) tools are \emph{interactive debugging tools} that operate passively without generating any new packets unlike the ATPG tool. The ndb tool \cite{handigol2012debugger}---the analogue of gdb debugger for programming---is like a network-wide path-aware tcpdump that builds packet histories which can be exploited by network analysis applications to verify the policy compliance of network data plane behavior.

\vspace{2mm}
\subsubsection{Control Plane Verification}
\label{sec:controlplaneverification}

Various projects have aimed at verification of the control plane functionality of SDNs. In the SDN architecture, it is envisioned that network programs will run as SDN applications on top of a northbound API exposed by SDN controller. This will allow SDN applications to leverage the services of the SDN controller, which will be responsible for managing the distributed state through a southbound API like OpenFlow, while the SDN application can focus on using the state for the task it wishes to perform. It is anticipated that this architecture will allow innovation to flourish and the development of numerous network based applications. In such an environment, it is necessary to ensure that we have tools available for testing and verifying such SDN applications. Canini et al. present their NICE framework for testing OpenFlow applications \cite{canini2012nice}. Kuzniar et al. have proposed another framework, named SOFT, for verifying OpenFLow switch interoperability. There also has been work on computationally verifying network programs in the Coq mechanical proof tool \cite{stewartcomputational}.

There also has been work on isolating fault inducing inputs to SDN control software \cite{scott2013did}, controller verification \cite{guha2013machine}, and ensuring per-packet and per-flow consistency of network updates \cite{reitblatt2012abstractions}. The problem of verifying a generic SDN controller---which in its general setting is Turing complete (e.g., NOX, Floodlight, etc.)---is undecidable.  Guha et al. have proposed a method of using for machine verification of network controllers \cite{guha2013machine}. FlowLog \cite{nelson2013balance} is a declarative, finite-state, language for programming SDN controllers that balances expressiveness and analysis and is amenable to model checking. In another model checking based work, Sethi et al. \cite{sethiabstractions} have proposed new data state and network state abstractions that can be used for model checking SDN controllers more efficiently. The Frenetic framework \cite{frenetic} incorporates features to help achieve per-packet and per-flow consistency during network updates \cite{reitblatt2012abstractions}. The safe update protocol proposed in \cite{reitblatt2012abstractions} builds upon approaches that use incremental recomputation (e.g., Anteater \cite{mai2011debugging}, VeriFlow \cite{khurshid2012veriflow}, etc.), which may have a transient stage in which the property to be verified may be violated, by ensuring that the property under check also holds during the transient stage.

\vspace{2mm}
\subsubsection{Network Debugging}
\label{sec:networkdebugging}

As mentioned before, networks are composed of both hardware and software components and are managed in many cases manually. Due to this reason, networks can fail in a variety of ways making the job of debugging and troubleshooting a network very complex. Traditionally, networking has a very primitive toolset for troubleshooting comprising few ad-hoc tools such as ping, traceroute, etc. usually complemented by the painstaking manual process of inspecting log files. Broadly speaking, debugging can take place either statically or dynamically. Static debugging---akin to compile-time checking---works by inspecting network configuration and settings through static analysis tools, model checking, SAT solvers, etc. Dynamic checking---similar to run-time checking---works by checking if the data plane is behaving as it should (techniques for data plane verification have earlier been discussed in section \ref{sec:dataplaneverification}). Dynamic checking can catch errors that arise from reasons other that erroneous configurations, e.g., it can help in the case of \emph{i)} hardware errors, \emph{ii)} link failure, \emph{iii)} congestion, \emph{iv)} intermittent problems, etc. Heller et al. have proposed systematic troubleshooting of SDNs by establishing equivalence of network views at different layers \cite{heller2013leveraging}. In particular, Heller et al. proposed comparing \emph{i)} actual network behavior vs. policy, \emph{ii)} the policy vs. device state, \emph{iii)} the device state vs. the hardware state, etc.  By comparing these diverse network views systematically, more efficient troubleshooting can be performed which will allow identification of faults and systematic tracking down root causes.

Handigol et al. \cite{handigol2012debugger} have proposed the ndb (network debugger) tool, analogous to the software debugging gdb tool, that aims to capture and reconstruct the sequence of events that leads to buggy behavior. In particular, it allows users to define a `network breakpoint' in the form of (header, switch) filter to identify the errant behavior, and then produces a packet backtrace, which includes historical information about the path taken by the packet as well as the state of the flow tables at each switch, to aid in troubleshooting of networks \cite{handigol2013using}. In a similar vein, Wundsam et al. \cite{wundsam2011ofrewind} have proposed the OFRewind framework which is useful for capturing and reproducing the sequence of problematic OpenFlow command sequence. In another work, Scott et al. have proposed using correspondence checking and simulation based causal inferencing to isolate and localize software faults in SDN \cite{scott2012and}. In networked systems, erroneous behavior can manifest itself due to the various issues related to distributed computing such as asynchrony, concurrency, and partial failures leading to time-consuming troubleshooting and considerable angst \cite{scott2013did}. Various debugging tools have been proposed for debugging general distributed systems: e.g., Pip \cite{reynolds2006pip}, etc., and automatic debugging techniques specific to SDN have been proposed in \cite{scottautomatic}.

\section{Open Issues and Future Work}
\label{sec:openissues}

The area of formal methods and verification is vast with various mature tools and techniques available. With networking being fundamentally important to all aspects of life including government, defence, industry, finance, etc., networks are in dire need of provably correct mechanisms. Notwithstanding the lack of any major breakthroughs made by formal methods in traditional networking, architectural support from SDN along with its clean abstractions provide a source of optimism for the future of formal methods in networking. The nascent field of network verification is wide open and is ripe for further exploration. In this section, we will point a few important open issues and highlight possible future work.

\subsection{Scalable Formal Verification For Large Networks}
\label{sec:scalable}

Advanced in technologies such as BDDs and SAT solvers have extended the state of the art considerably in recent years. However, more work needs to be done for current formal verification techniques to scale to large networks and to verify large software systems (such as network applications and protocol implementations). An approach that has been proposed in literature for scaling to large networks is to utilize incremental recomputation thereby avoiding the overhead of redoing expensive static calculations. For example, NetPlumber \cite{kazemian2012header} improves HSA \cite{kazemian2012header}, and Veriflow \cite{khurshid2012veriflow} improves Anteater \cite{mai2011debugging}, by supporting incremental computation. The incorporation of incremental recomputations techniques have allowed these tools to scale to reasonably large networks. In recent work, Yang and Lam have proposed an efficient real-time verifier of network properties using atomic predicates \cite{yangreal}. More work is needed in this area to exploit these recent works so that network verification for large networks can become both practical and efficient.

\subsection{Automated Synthesis}
\label{sec:autosynthesis}

Synthesis which promises to automatically derive implementations from specifications is an extremely important future goal that can improve programmer productivity. The problem of automated synthesis is at the frontier of verification research today \cite{lopesnetwork}. Some important works in this regard include synthesis of network updates \cite{noyestoward}, synthesis of network controllers \cite{wangautomated2013}, synthesis of finite state controllers from temporal logic specifications, and synthesis of programs from examples by exploiting domain specific knowledge, etc. \cite{alurtheory}. In the context of networking, more work needs to be done so that subsystems such as protocols, configurations, hardware may be synthesized through a high-level formal specification only in a user-friendly manner.

\subsection{Selection of the Right Formal Method for the Task}
\label{sec:rightmethod}

As highlighted in this work, there is a vast amount of work that has been done in the field of formal methods. There are various logics, notations, technologies and tools available, each making its own claim of superiority, that may be utilized. Many of the claims are valid in that certain tools do certain excel in niche areas; however, each tool has its disadvantages as well. As Keshav pointed out in \cite{keshavmodeling}, the choice of the most appropriate tool is certainly not trivial even for an established researcher, let alone for a graduate student. It is important to use the most appropriate specification language for the task, as noted in the 10 commandments stated in \cite{bowen1995ten}. With research in network verification recently starting to flourish, it is important to determine the right tools for various verification tasks in network verification. Two tools that are immediately useful for a networking researcher are Alloy and SPIN: a practical comparison of these two tools is presented in \cite{zavepractical}.

\subsection{Specialized Network Verification Tools}
\label{sec:specializedtools}

In contrast to sophisticated well-honed design automation tools that are available for general hardware\footnote{The electronic design automation (EDA) industry in hardware design is a big market catering to a multi-billion dollar industry.} and software industries, networking industry has almost no rigorous tools for verification. The vision of building a network CAD was articulated by McKeown. Encouragingly, as the SDN architecture is becoming mainstream, there is renewed interest in building specialized tools that will allow automated debugging, verification, and analysis. Some important issues that need to be addressed before such a vision can be realized are \cite{heller2013leveraging}: \emph{i)} incorporating program semantics into network troubleshooting tools; \emph{ii)} improved techniques for checking invariants; \emph{iii)} development of new abstractions, especially in the SDN context, to facilitate troubleshooting.

\subsection{Verification for Concurrent and Parallel Programming}
\label{sec:specializedtools}

With the emergence of data centers and cloud computing, the programming world is undergoing a silent revolution with a growing trend towards parallel programming. Although, there are various approaches that have been proposed to support verification of concurrent programs, more research needs to be done to propose new clean simplified abstractions for building verified concurrently executing programs that can exploit modern multi-core and multi-processor architectures, and parallel programming style suited to data centers and cloud computing.

\section{Conclusions}
\label{sec:conclusions}

We are in an exciting time in the networking world with recent innovations such as software defined networking and cloud computing fundamentally altering the landscape of the networking world. Keeping in mind the criticality of the Internet infrastructure, assuring the correct behavior of various subsystems of the Internet has become essential.  There is great interest in applying the vast amount of work that has been done in the community of formal methods and verification to networks. The work in formal methods draws upon many diverse fields such as logic, theoretical computer science, programming languages, mathematics, etc., and hence appears daunting to a non-specialist. In this work, we present a detailed tutorial on the various methods and techniques used in formal methods and verification while providing necessary background and references to important works. We also present a detailed survey of the application of formal methods in the networking context. We have also identified some important research directions that can be pursued in future work.

\bibliographystyle{ieeetr}
\bibliography{formalmethods}

\begin{IEEEbiography}[{\includegraphics[width=1in,height=1.25in,clip,keepaspectratio]{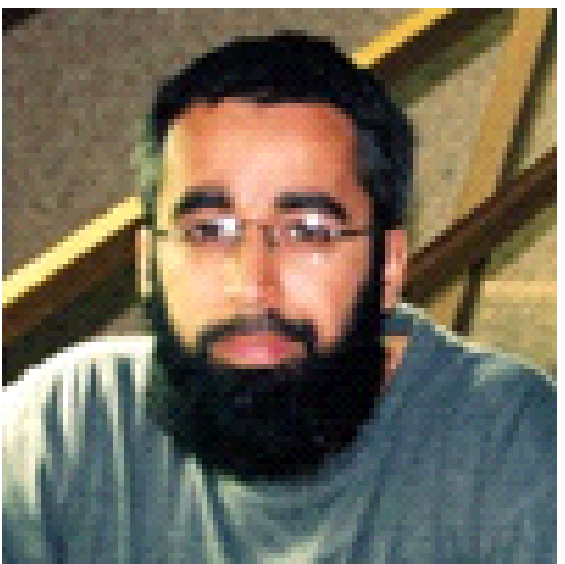}}]{Junaid Qadir}
He is an Assistant Professor at the School of Electrical Engineering and Computer Sciences (SEECS), National University of Sciences and Technology (NUST), Pakistan. He is also the Director of the Cognet Lab at SEECS. He completed his BS in Electrical Engineering from UET, Lahore, Pakistan and his PhD from University of New South Wales, Australia in 2008. His research interests include networking/ algorithmic issues in cognitive radio networks, wireless networks, and software-defined networks.
\end{IEEEbiography}

\begin{IEEEbiography}[{\includegraphics[width=1in,height=1.25in,clip,keepaspectratio]{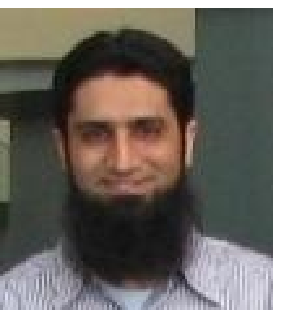}}]{Osman Hasan}
He is an Assistant Professor at the School of Electrical Engineering and Computer Sciences (SEECS), National University of Sciences and Technology (NUST), Pakistan. He is the lab director of System Analysis and Verification (SAVe) Lab at NUST SEECS.  His main research interests include formal verification, interactive theorem proving and higher-order logic. Prior to joining NUST SEECS, he did his PhD and post-doctoral fellowship from the hardware verification group (HVG) at Concordia University, Montreal, Canada.
\end{IEEEbiography}

\end{document}